\documentclass[lettersize,journal]{IEEEtran}
\usepackage{amsmath,amsfonts,amssymb}
\usepackage{algorithmic}
\usepackage{algorithm}
\usepackage{array}
\usepackage[caption=false,font=normalsize,labelfont=sf,textfont=sf]{subfig}
\usepackage{textcomp}
\usepackage{stfloats}
\usepackage{url}
\usepackage{verbatim}
\usepackage{graphicx}
\usepackage{cite}
\usepackage[subtle]{savetrees}
\usepackage[short]{optidef}
\usepackage{epstopdf}
\allowdisplaybreaks[4]
\usepackage{xcolor}

\begin{document}

\def\x{{\mathbf x}}
\def\L{{\cal L}}

\title{Waveform Design for Wireless Power Transfer with Power Amplifier and Energy Harvester Non-Linearities}
\author{Yumeng Zhang and \IEEEauthorblockN{Bruno Clerckx, \IEEEmembership{Fellow,~IEEE,} \thanks{The authors are with the Department of Electrical and Electronic Engineering, Imperial College London, London SW7 2AZ, U.K. (e-mail:
b.clerckx@imperial.ac.uk, yumeng.zhang19@imperial.ac.uk). B. Clerckx is also with Silicon Austria Labs (SAL), Graz A-8010, Austria.} \thanks{Part of section IV-C in this paper is published at the 2022 International Conference on Acoustics, Speech, and Signal Processing in \cite{9746380}. Compared with the conference version, this paper: 1) proposes a new  model that addresses the drawbacks of the model in \cite{9746380}, i.e., the resultant input  {waveform} at the HPA incurs infinite bandwidth and the optimization in each iteration involves an integral inequality constraint with high computational complexity; 2) provides details and analysis of the resultant waveform in \cite{9746380}; 3) gives insight into the  {conflicting} effect of HPA's and EH's non-linearities on both the power harvesting performance and the waveform  {shapes}.}}}
\maketitle

\begin{abstract}
Waveform optimization has shown its great potential to boost the performance of far-field wireless power transfer (WPT). Current research has optimized transmit waveform, adaptive to channel state information, to maximize the harvested power in WPT while accounting for the energy harvester (EH)’s non-linearity. However, the existing transmit waveform design disregards the non-linear high power amplifiers (HPA) at the transmitter. Driven by this, this paper optimizes a multi-carrier waveform at the input of HPA to maximize the harvested DC power considering both HPA’s and EH’s non-linearities. Two optimization models are formulated based on whether the frequencies of the  {input} waveform are concentrated within the transmit pass band or not. Analysis and simulations show that, while EH’s non-linearity boosts the power harvesting performance, HPA’s non-linearity degrades the harvested power. Hence, the optimal waveform shifts from multi-carrier that exploits EH’s non-linearity to single-carrier that reduces HPA’s detrimental non-linear distortion as the operational regime of WPT becomes more sensitive to HPA’s non-linearity and less sensitive to EH’s non-linearity (and inversely). Simultaneously, operating towards HPA’s non-linear regime by increasing the input signal power benefits the harvested power since HPA’s DC power supply is better exploited, whereas the end-to-end power transfer efficiency  might decrease because of  {HPA's} increasing non-linear degradation. Throughout the simulations, the proposed waveforms show significant gain over those not accounting for HPA’s non-linearity, especially in frequency-flat channels. We also compare the two proposed waveforms and show that the severity of HPA’s non-linearity dictates which of the two proposed waveforms is more beneficial.
\end{abstract}

\begin{IEEEkeywords}
Waveform design, power harvesting, non-linearities, power amplifier, wireless power transmission
\end{IEEEkeywords}
\section{Introduction}
\label{sec:intro}
Future networks are expected to cope with the emergence of trillions of low-power sensors for applications in the Internet of Things (IoT). However, the explosion in the number of these low-power devices is challenging traditional battery power supply  {that needs periodic replacement or wired power supply}. Nevertheless,  {thanks to the advances in efficient electronics and computation architectures, the power consumption of these remote devices reduces dramatically, which  gives rise to far-field wireless power transfer (WPT) as an alternative source of power} \cite{mustafa2021joint,ijemaru2021mobile}.

Far-field WPT generates and transmits suitable RF signals that propagate over the air before being captured and rectified into DC current via rectenna circuits at the receivers.  The availability of RF signals as a power source releases WPT from batteries or wires, which makes it more sustainable, flexible, and controllable.  Since the discovery of WPT's promising potential to address the powering bottlenecks in IoT,  scientists have been making every effort to boost the  {end-to-end power transfer efficiency (PTE)} of WPT on several fronts, namely, circuits design \cite{suh2002high,1556784}, RF signals design \cite{clerckx2021wireless,BrunoToward},  {reconfigurable intelligent reflective surfaces (RIS) and} materials \cite{dickinson1975radiated,clerckx2021wireless,feng2021waveform}.

Signals and waveform{s} play a crucial role in the  {end-to-end} PTE of WPT,  {which is the product of the DC-to-RF conversion efficiency at the transmitter, the over-the-air RF-to-RF conversion efficiency, and the RF-to-DC conversion efficiency at  the receiver \cite{clerckx2021wireless}.}  {As part of the overall PTE, the RF-to-RF and the RF-to-DC conversion efficiency} have been comprehensively experimented given different types of signals and  {energy harvester (EH)/rectenna} structures \cite{collado2012improving, Clerckx2016Waveform,boaventura2015boosting}. Experimental research verifies the advantages of    {receiving} signals featuring a high peak-to-average-power ratio (PAPR) regarding the RF-to-DC  {conversion efficiency}, but without consideration of channel state information  (CSI) \cite{Trotter2009Power,boaventura2015boosting}.  A noteworthy contribution later is in \cite{Clerckx2016Waveform} where  {a multi-carrier transmit waveform emitted at the transmitter was}  theoretically designed, adaptive to CSI and rectenna's non-linear characteristics, to maximize the  { DC current at the rectenna's output, so that the product of the RF-to-RF and the RF-to-DC conversion efficiency was maximized}.  Specifically, \cite{Clerckx2016Waveform} characterized the non-linear rectenna via Taylor expansion and then expressed the generated DC power as a non-linear function of the transmitted waveform.  {Waveform optimization based on this non-linear rectenna model} has shown a significant gain in the power harvesting performance over those assuming linear rectennas \cite{HuangLarge2017,abeywickrama2021refined,BrunoToward}.

The model in \cite{Clerckx2016Waveform}  gives rise to further flourishment in the field of signal design in far-field WPT. A corresponding low-complexity algorithm was developed in \cite{ClerckxA}, and the efficiency of these waveforms was validated through  prototyping and experimentation in \cite{KimSignal}. Moreover, the model has been extended to general scenarios such as limited-feedback\cite{Huang1}, large-scale \cite{HuangLarge2017}, multi-input-multi-output \cite{shen2020beamforming}, multi-user \cite{HuangLarge2017,abeywickrama2021refined} and opportunistic/fair-scheduling \cite{kim2020opportunistic,8476162}. For further improvement on  the joint RF-to-RF and RF-to-DC conversion efficiency, hybrid beamforming and RIS have also been exploited \cite{shen2021joint,feng2021waveform,zhao2021irs}. Besides the performance advancement, WPT has also found applications in integrated systems such as wireless information and power transfer (WIPT) \cite{clerckx2017wireless} and wireless powered backscatter communications \cite{clerckx2017wirelessly}.

However, the above waveforms were performed without much consideration of the  {DC-to-RF conversion efficiency as part of the overall PTE, i.e., they neglected the} transmitter's architecture,  especially the most non-linear and power-consuming component, namely the high power amplifier (HPA) at the transmitter.  Indeed, recent research has experimentally verified and characterized the performance degradation caused by non-linear HPAs in WPT/WIPT systems \cite{2020WIPTNON,jang2020novel,krikidis2020information,park2020performance,chen2021performance}.  Effort has also been made to explore the beneficial signals that  trade off between  HPA's and EH's non-linearities in WPT systems\cite{9746380}.
\cite{ayir2021waveforms} claimed the advantage of  { single-carrier} QPSK signals over uniformly power-allocated multi-carrier signals in the presence of HPA's and EH's non-linearities, but was based on a simplified platform compared with \cite{KimSignal} and ignored the performance gain achievable from the CSI knowledge. An interesting observation is in \cite{clerckx2021wireless}  {where the input constellation of WIPT was optimized using machine learning techniques. In \cite{clerckx2021wireless}, WIPT's optimal input constellation shifted from forming a single high-amplitude WPT mass point (with other low-amplitude WIT mass points) given a linear HPA to forming two WPT mass points with relatively lower amplitudes given a non-linear HPA, to combat HPA's non-linear degradation}.   {The input distribution was also optimized in WIPT with HPA's and EH's  non-linearities  in \cite{9774343} }. However, the paper  was based on a single-carrier assumption, getting rid of the problems of inter-carrier interference and frequency leakage from  HPA's and EH's  non-linearities in a multi-carrier scenario.  This drives our interest to explore the fundamental question of   {how the optimal multi-carrier waveform changes when HPA's non-linearity is also considered in WPT}.

To combat HPA's non-linear effect at the waveform-design level, mainly two lines of methods have been proposed, namely designing signals less susceptible to HPA's non-linearity and by means of digital pre-distortion (DPD). The former method decreases the input signals' exposure to HPA's non-linear region, such as PAPR reduction \cite{kryszkiewicz2018amplifier}, distortion power minimization \cite{kryszkiewicz2018amplifier} and out-of-band leakage power reduction \cite{goutay2021end}. Indeed, PAPR reduction has been introduced as a transmit waveform constraint in WPT in \cite{Clerckx2016Waveform}. However, this class of methods might be less efficient in the system-level design since they either neglect channels and receiver architectures \cite{kryszkiewicz2018amplifier,goutay2021end} or dismiss HPA's transfer characteristics \cite{Clerckx2016Waveform}. In contrast, DPD pre-distorts the desired input signal according to HPA's transfer characteristics to linearize the transfer function of the joint pre-distorter-and-HPA structure or, if not achievable, to minimize the non-linear effect of the HPA on system performance\cite{fu2014frequency}. In the latter case, the signal is designed to compensate for HPA's non-linearity and also to adapt to system characteristics.

Following the above analysis, this paper explores designing the  {predistorted} input waveform at the HPA to boost the power harvesting performance in a practical  {multi-carrier} WPT system accounting for both HPA's and rectenna's non-linearities.  {Notice that although the problem has been firstly handled in \cite{9746380}, the method therein is to optimize the transmit signal and hence suffers from a corresponding input signal at the HPA with unlimited bandwidth, whose effect is left uncovered. Also, while the paper  {shows} the performance gain by considering HPA's non-linearity, it lacks insight into how the optimal waveform is affected by HPA's non-linearity and how it trades off between HPA's and EH's non-linearities, which should be a significant feature to inspire any future wireless powered systems.} Overall, the contributions of the paper are summarized as follows:

(1) First, we model a practical WPT system accounting for both  transmitter's and receiver's non-linearities as in Fig. \ref{Fig_whole_structure}. The transmitter is composed of a non-linear HPA and a  {band pass filter (BPF)} without loss of generality. On this basis, we develop the relationship between the input waveform of HPA $x^{\text{in}}(t)$ ({namely} the input  waveform), the  waveform after HPA $x^{\text{HPA}}(t)$ and the transmit  waveform at the antenna $x^{\text{tr}}(t)$ ({namely} the transmit  waveform).  {The transmit waveform, $x^{\text{tr}}(t)$, after propagation over the air, is picked up and transferred into DC power by a non-linear rectenna as in \cite{Clerckx2016Waveform}}. Based on the transmitter's and rectenna's model, we formulate an end-to-end waveform optimization problem that maximizes the generated DC power given power constraints both at the HPA's input  {(the input power constraint)} and at the transmit antennas  {(the transmit power constraint)}. This contrasts with all prior works on waveform optimization  {in} \cite{Clerckx2016Waveform,ClerckxA,clerckx2021wireless,HuangLarge2017} that consider non-linearity at the rectenna side only.

(2) Second, we formulate and solve a first optimization model where the input waveform  shares the same frequencies as those of the transmit waveform (within the transmit pass band), abbreviated as Model I. In the problem formulation,  {the harvested power (objective function) and the transmit power constraint are expressed as a function of the transmit waveform instead of the input waveform to avoid high complexity, by introducing the transmit waveform as auxiliary variables. The relationship between the input waveform and the transmit waveform is clarified through multiple non-linear equalities.} We adopt successive convex programming (SCP) to approximate the objective function first and then use sequential quadratic programming (SQP) to handle the non-linear constraints, abbreviated as SCP-SQP.  {Model I tackles the problems in \cite{9746380} (particularly the input waveform with unlimited bandwidth) by realizing an end-to-end optimization where the input waveform is optimized directly.}

(3) Third, we formulate and solve a second optimization model where the out-of-band frequencies of the transmit frequencies are utilized additionally when designing the input waveform, abbreviated as Model II. The aim is to concentrate the frequencies of the waveform after the nonlinear HPA within the pass band of the BPF in Fig. \ref{Fig_whole_structure} (or the transmit pass band) to avoid power loss at BPF. This assumption allows the input power constraint to be re-formulated as a function of the transmit waveform, making the whole optimization  {with respect to} the transmit waveform only (instead of both the input and transmit waveforms as in Model I). Consequently, this halves the number of variables in Model I and avoids numerous non-linear equality constraints, making the program more stable in large-scale problems. The resulting problem is solved by SCP and the interior-point (IP) method, abbreviated as SCP-IP. {Model II corresponds to the conference paper in \cite{9746380},  but is complemented in this paper by providing an approximated input signal whose frequency components out of the  {transmit pass band}  (or an extended bandwidth dependent on HPA's non-linearity) are cut off. }

(4) Fourth, we provide simulation results to verify the performance gain of waveforms that consider both HPA's and EH's non-linearities over those that ignore HPA's non-linearity. We also demonstrate that HPA's non-linearity, in sharp contrast with EH's non-linearity, degrades the harvested power in WPT significantly, especially in frequency-flat channels.  Simulations also show that operating towards the non-linear regime of HPA exploits its DC power supply more and gives larger harvested power and higher end-to-end PTE. However, if further increasing the input signal power to where the HPA saturates, the harvested power saturates and the end-to-end PTE decreases correspondingly. To compensate for the power loss from HPA's non-linearity, the proposed waveforms show a tendency to allocate power to fewer sub-carriers as the HPA experiences more non-linearity. {Given this general tendency, the performance of Model I and Model II are similar in most simulations, while scenarios are also explored to make the waveform from Model I (or Model II) slightly more advantageous than the other by adapting the HPA's parameters and operating region.}

\textit{Organization:} Section II introduces the system  model and section III models the non-linear HPA and rectenna. Section IV addresses the waveform  optimization given two different assumptions on the frequencies of the input signal at HPA. Section V evaluates the performance and section VI concludes the work.

\textit{Notations:} Matrices and vectors are respectively denoted in form of bold upper case and bold lower case. Complex symbols are characterized by a tilde superscript, i.e., $\widetilde{x}$. $\mathfrak{R} \left\lbrace\widetilde{x}\right\rbrace  $ and $\mathfrak{I} \left\lbrace\widetilde{x}\right\rbrace  $ denote the real and imaginary part of the complex number $\widetilde{x}$ respectively, and are referred to as $\overline{x}$ and $\widehat{x}$. ${x}$ represents the amplitude of the complex number $\widetilde{x}$  {and $\measuredangle \widetilde{x}$ represents the phase of $\widetilde{x}$}. $\left\lbrace    x_n\right\rbrace  $ represents the set that collects all the related $x_n$.   {$\circledast_N$ means the circular convolution of length $N$. For integer $a$ and $b$, $(a)_{b}$ means $a$ modulo $b$. $\mathcal{E}\left\lbrace    x(t)\right\rbrace  $ represents taking average of $x(t)$ over time.}

$\mathcal{F}\left\lbrace x(t)\right\rbrace$  {conducts Fourier Transform (FT)} FT on $x(t)$ and $\mathcal{F}\left\lbrace x(t)\right\rbrace|_{f=f_0}$ takes the spectrum at frequency $f_0$. For discrete Fourier Transform (DFT),  {define $n$ as the index across the frequency domain and $k$ as the index across the time domain. The $n^{\text{th}}$ entry of a $K$-point DFT on series $x[k]$ is given by:}
\begin{align}
x[n]=\text{DFT}\left\lbrace     x[k]\right\rbrace_{(n)}=\sum_{k=0}^{K-1} x[k]e^{\frac{-2j\pi kn}{K}}.
\end{align}

\section{WPT System Model}\label{section_WPT_system_model}

\begin{figure*}[t]
\centering
\includegraphics[scale=0.85]{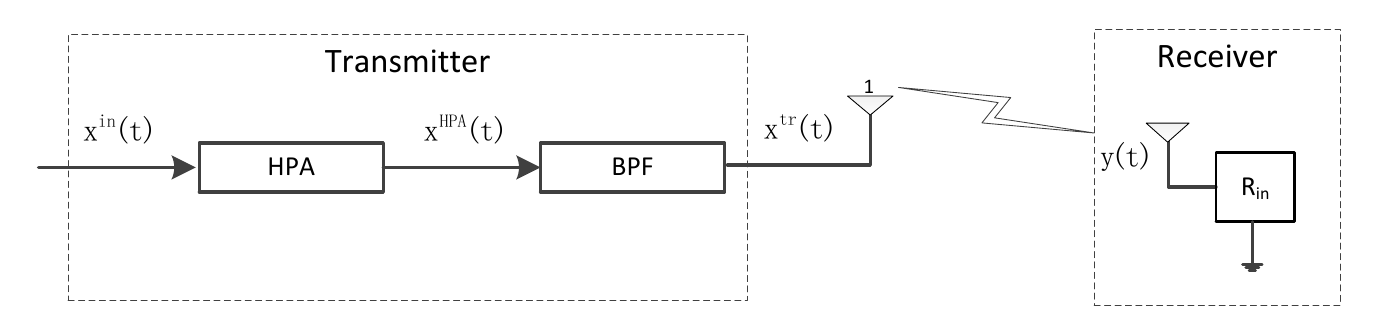}
\caption{The system model of WPT with non-linear HPA and EH.}
\label{Fig_whole_structure}
\end{figure*}
In this section, we introduce a practical WPT architecture with a non-linear HPA at the transmitter and a non-linear EH at the receiver. At the transmitter, the generated multi-carrier signal is first amplified and then filtered before being transmitted. The RF signal then propagates through Rayleigh fading channels and is picked up by the EH for the purpose of power harvesting.  The functioning of different building blocks will be clarified throughout the paper.

We consider a single-user single-input-single-output (SISO) multi-carrier WPT system as depicted in Fig. \ref{Fig_whole_structure}. The non-linear analysis of the transmitter starts from the input of the non-linear HPA at the RF chain. For simplicity, we denote the complex  {input signal at the HPA as $\widetilde{x}^{\text{in}}(t)$, the complex signal at the output of the HPA as $\widetilde{x}^{\text{HPA}}(t)$, and the complex transmit signal as $\widetilde{x}^{\text{tr}}(t)$.}  We consider a multi-carrier scenario where $N$ evenly frequency-spaced sub-carriers are finally transmitted. The frequency of the $n^{\text{th}}$ $(n=0,~\cdots,~N-1)$ transmitted sub-carrier of $\widetilde{x}^{\text{tr}}(t)$ is $f_n=f_0+(n-1)\Delta_f$ with $f_0$ being the lowest sub-carrier frequency and $\Delta_f$ being the frequency spacing.

At the transmitter, the complex  {input} signal at the  {HPA} is:
\begin{align}
\label{eq_input_signal_complex1}
\widetilde{x}^{\text{in}}(t)&={x}^{\text{in}}(t)e^{j\measuredangle \widetilde{x}^{\text{in}}(t)}=\sum_{n\in \kappa^{\text{in}}}\widetilde{w}^{\text{in}}_{n}e^{j2\pi f_nt}=\sum_{n\in \kappa^{\text{in}}} \left(\overline{w}^{\text{in}}_{n}+j\widehat{w}^{\text{in}}_{n}\right)e^{j2\pi f_nt},
\end{align}
where ${x}^{\text{in}}(t)=|\widetilde{x}^{\text{in}}(t)|$ is the amplitude of $\widetilde{x}^{\text{in}}(t)$ and $\measuredangle \widetilde{x}^{\text{in}}(t)$ is the phase of $\widetilde{x}^{\text{in}}(t)$. $\widetilde{w}^{\text{in}}_{n}=\overline{w}^{\text{in}}_{n}+j\widehat{w}^{\text{in}}_{n}$ denotes the complex weight of the $n^{\text{th}}$ sub-carrier  { of} $ \widetilde{x}^{\text{in}}(t)$, with $\overline{w}^{\text{in}}_{n}$ and $\widehat{w}^{\text{in}}_{n}$ being the corresponding real and imaginary part respectively. $\kappa^{\text{in}}$ is the integer set which indicates the frequency range that $\widetilde{x}^{\text{in}}(t)$ involves \footnote{{If $\kappa^{\text{in}}=[0,~\cdots,~N-1]$, $\widetilde{x}^{\text{in}}(t)$ will have the same activated frequencies with the transmit signal $\widetilde{x}^{\text{tr}}(t)$. In this case, the signal after HPA, $\widetilde{x}^{\text{HPA}}(t)$, will incur out-of band frequencies (at the multiples of $\Delta_f$ due to the HPA's non-linearity) which are going to be filtered out by the BPF in Fig. \ref{Fig_whole_structure}. This corresponds to the first model (Model I) in the paper. }}$^{,}$\footnote{Otherwise,  {if assuming the input signal with unlimited bandwidth, i.e., $\kappa^{\text{in}}=[-f_0/\Delta_f+1,~\cdots,~\infty]$, the input signal $\widetilde{x}^{\text{in}}(t)$ can be designed to concentrate the frequency components of $\widetilde{x}^{\text{HPA}}(t)$ within the transmit pass band}. In this case, the signal after the non-linear HPA can pass the BPF in a lossless way, i.e., $\widetilde{x}^{\text{HPA}}(t)=\widetilde{x}^{\text{tr}}(t)$. This corresponds to the second model (Model II) in the paper.}.

Similarly, the complex signal at the output of the HPA is:
\begin{align}
\label{eq_signal_HPA_complex}
\widetilde{x}^{\text{HPA}}(t)&=\sum_{n\in \kappa^{\text{HPA}}}\widetilde{w}^{\text{HPA}}_{n}e^{j2\pi f_nt}
=\sum_{n \in\kappa^{\text{HPA}}} \left(\overline{w}^{\text{HPA}}_{n}+j\widehat{w}^{\text{HPA}}_{n}\right)e^{j2\pi f_nt},
\end{align}
where $\widetilde{w}^{\text{HPA}}_{n}=\overline{w}^{\text{HPA}}_{n}+j\widehat{w}^{\text{HPA}}_{n}$ denotes the complex weight of the $n^{\text{th}}$ sub-carrier  { of} $ \widetilde{x}^{\text{HPA}}(t)$, with $\overline{w}^{\text{HPA}}_{n}$ and $\widehat{w}^{\text{HPA}}_{n}$ being the corresponding real and imaginary part respectively {($\overline{w}^{\text{HPA}}_n/\widehat{w}^{\text{HPA}}_n$ for $\forall$ $n$ is a function of $\left\lbrace\overline{w}^{\text{in}}_n,~\widehat{w}^{\text{in}}_n\right\rbrace$,  and can be viewed as a simplified form of $\overline{w}^{\text{HPA}}_n\left(\left\lbrace\overline{w}^{\text{in}}_n,~\widehat{w}^{\text{in}}_n\right\rbrace\right)/\widehat{w}^{\text{HPA}}_n\left(\left\lbrace\overline{w}^{\text{in}}_n,~\widehat{w}^{\text{in}}_n\right\rbrace\right)$, whose relationship is characterized later in Section III.).} $\kappa^{\text{HPA}}$ represents the frequency set that  $\widetilde{x}^{\text{HPA}}(t)$ involves.

The BPF then removes the frequencies out of the transmit passband $[f_0,~f_0+(N-1)\Delta_f]$ , which gives the real transmit signal as:
\begin{align}
\label{eq_signal_tr_complex}
\nonumber\overline{x}^{\text{tr}}(t)&=\mathfrak{R} \left\lbrace\widetilde{x}^{\text{tr}}(t)\right\rbrace=\mathfrak{R} \left\lbrace\sum_{n =0}^{N-1}\widetilde{w}^{\text{tr}}_{n}e^{j2\pi f_nt}\right\rbrace\\
&=\mathfrak{R}\left\lbrace \sum_{n =0}^{N-1} \left(\overline{w}^{\text{tr}}_{n}+j\widehat{w}^{\text{tr}}_{n}\right)e^{j2\pi f_nt}\right\rbrace,
\end{align}
where $\widetilde{x}^{\text{tr}}(t)$ is the complex transmit signal.  $\widetilde{w}^{\text{tr}}_{n}=\overline{w}^{\text{tr}}_{n}+j\widehat{w}^{\text{tr}}_{n}$ denotes the complex weight of the $n^{\text{th}}$ sub-carrier  { of} $ \widetilde{x}^{\text{tr}}(t)$, with $\overline{w}^{\text{tr}}_{n}$ and $\widehat{w}^{\text{tr}}_{n}$ being the corresponding real and imaginary part respectively {($\overline{w}^{\text{tr}}_n/\widehat{w}^{\text{tr}}_n$ for $\forall$ $n$ is a function of $\left\lbrace\overline{w}^{\text{in}}_n,~\widehat{w}^{\text{in}}_n\right\rbrace$,  and can be viewed as a simplified form of $\overline{w}^{\text{tr}}_n\left(\left\lbrace\overline{w}^{\text{in}}_n,~\widehat{w}^{\text{in}}_n\right\rbrace\right)/\widehat{w}^{\text{tr}}_n\left(\left\lbrace\overline{w}^{\text{in}}_n,~\widehat{w}^{\text{in}}_n\right\rbrace\right).)$}.

The transmitted signal $\overline{x}^{\text{tr}}(t)$ then propagates through a multipath channel with $L$ paths. Assume that the signal received from the $l^{\text{th}}$ $(l=1,~\cdots,~L)$ path is:
\begin{align}
\label{eq_WPT_received_signal}
\overline{y}_l(t)=\mathfrak{R}\left\lbrace\sum_{n=0}^{N-1}\alpha_{l}\widetilde{w}^{\text{tr}}_{n}e^{j(2\pi f_n(t-\tau_l)+\zeta_{l})}\right\rbrace,
\end{align}
where $\alpha_{l}$ denotes the amplitude suppression degree at the $l^{\text{th}}$ path and $\tau_l$ denotes the path delay, assuming a narrowband balanced array here. $\zeta_{l}$ is the phase shift propagating through the $l^{\text{th}}$ path. Then, the total  {real received} signal is the sum over Eq. \eqref{eq_WPT_received_signal}:
\begin{align}
\label{eq_WPT_received_signal2}
\nonumber\overline{y}(t)&=\sum_{l=1}^L \overline{y}_l(t)=\mathfrak{R}\left\lbrace\sum_{n=0}^{N-1} \left[ \sum_{l=1}^L \alpha_{l}e^{j2\pi (-f_n\tau_l+\zeta_{l})}\right] \widetilde{w}^{\text{tr}}_{n}e^{j2\pi f_nt}\right\rbrace\\
&\overset{\Delta}{=}\mathfrak{R}\left\lbrace\sum_{n=0}^{N-1}\widetilde{h}_{n}\widetilde{w}^{\text{tr}}_{n}e^{j2\pi f_nt}\right\rbrace,
\end{align}
where $\widetilde{h}_{n}=\sum_{l=1}^L \alpha_{l}e^{j2\pi (-f_n\tau_l+\zeta_{n,l})}$ denotes the equivalent complex channel that the $n^{\text{th}}$ sub-carrier experiences during transmission.

\section{HPA and Rectenna Analytical Models}
This section models the HPA's and rectenna's transfer characteristics in WPT.  We develop the relationship between the signals at each stage, and finally characterize the harvested power explicitly.

\subsection{HPA Model}
\subsubsection{HPA model characterization}
The input signal $\widetilde{x}^{\text{in}}(t)$ is amplified and filtered before being transmitted. Generally, a practical HPA will exert non-linear amplitude distortion and phase shift on its input signal. The HPA's distorted effect on both amplitude and phase are normally characterized by functions with respect to the amplitude of the input signal. Mathematically, the transfer characteristic of a practical HPA can be interpreted by:
\begin{align}
\label{eq_HPA}
f_{\text{HPA}}(\widetilde{x})=\mathcal{A}(x)e^{j[\measuredangle \widetilde{x}+\mathcal{\phi}(x)]},
\end{align}
where $\widetilde{x}=xe^{j\measuredangle \widetilde{x}}$ with $x$ being the amplitude of $\widetilde{x}$ and $\measuredangle \widetilde{x}$ being the phase of  $\widetilde{x}$. Function $\mathcal{A}(x)$ represents the amplitude distortion of  {the} HPA, and function $\mathcal{\phi}(x)$  characterizes the phase shift  {of} the HPA.

\begin{figure}[t]
\centering
\includegraphics[width=0.45\textwidth]{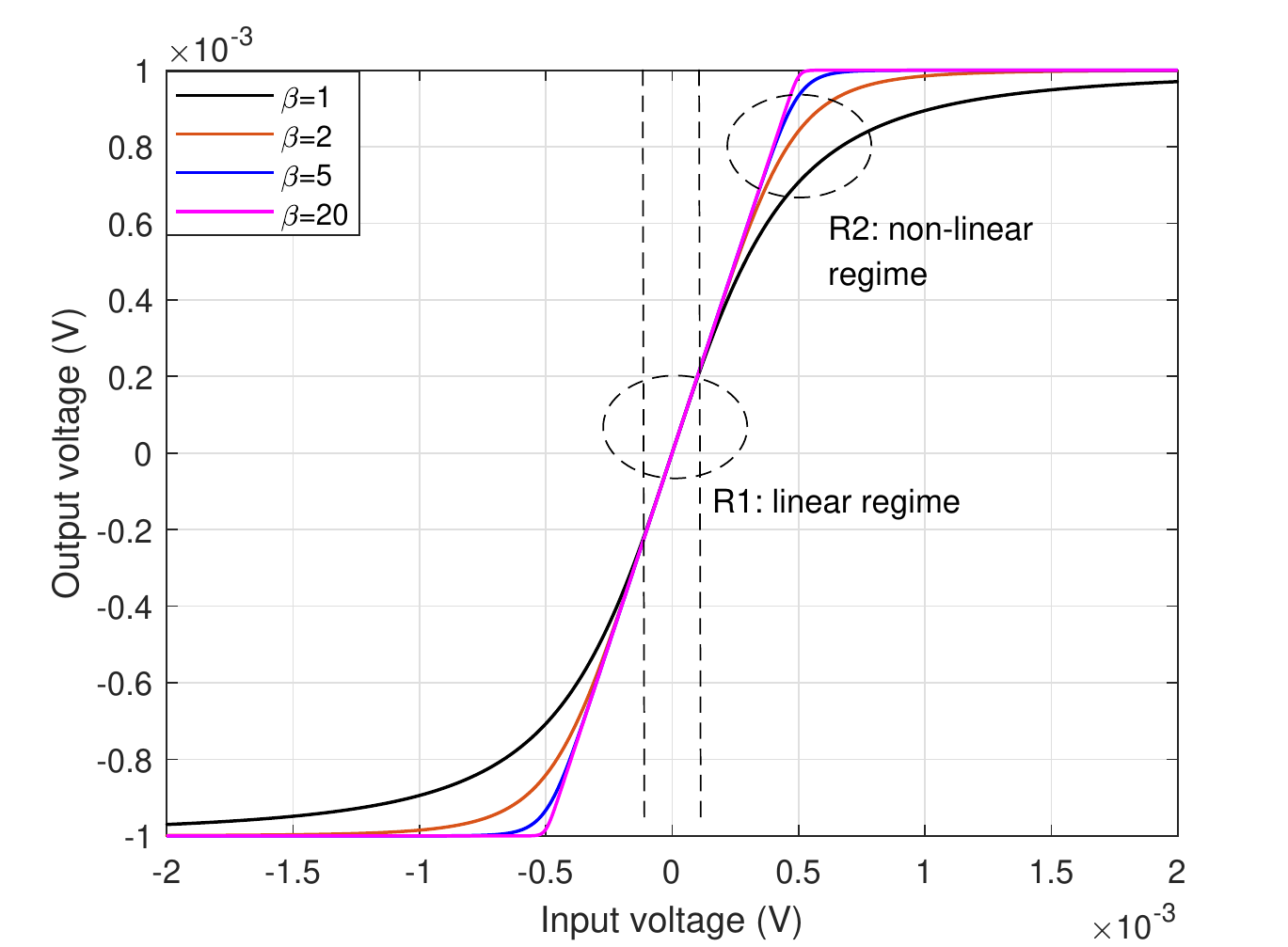}
\caption{\footnotesize SSPA's input-output voltage characteristics. $A_s=1$ mV; $G=2$ and $\beta=\left\lbrace    1,~2,~5,~20\right\rbrace  $. $R_1$ is the linear regime (small input signal); $R_2$ is the non-linear regime.}
\label{fig_SSPA_model}
\end{figure}

In this paper, we focus on a classic solid-state-power amplifier (SSPA) modelled in \cite{rapp1991effects}:
\begin{align}
\label{eq_HPA2_amp}
\mathcal{A}(x)&=\frac{{Gx}}{\left[1+\left(\frac{Gx}{A_s}\right)^{2\beta}\right]^{\frac{1}{2\beta}}},\\
\label{eq_HPA2_phase}\mathcal{\phi}(x)&=0,
\end{align}
where $G$ is the small-signal amplifier gain of SSPA, $A_s$ is the saturation voltage of SSPA, and $\beta$ (positive) is the smoothing parameter of SSPA. Eq. \eqref{eq_HPA2_phase} indicates that the SSPA does not have phase shift. Features of SSPA are clarified in Fig. \ref{fig_SSPA_model}, where the small signal input is offered an amplifier gain of $G=2$ approximately while the large signal input is shown to be suppressed below the saturation voltage of SSPA $(A_s)$. The larger the smoothing parameter $\beta$, the higher the linearity SSPA exhibits below its saturation voltage.

Combing Eq. \eqref{eq_input_signal_complex1} and Eq. \eqref{eq_HPA}, the complex signal at the output of the HPA (SSPA) becomes:
\begin{equation}
\label{eq_HPA_model}
\widetilde{x}^{\text{HPA}}(t)=f_{\text{HPA}}\left(\widetilde{x}^{\text{in}}(t)\right)=\frac{G\widetilde{x}^{\text{in}}(t)}{\left[1+\left(\frac{Gx^{\text{in}}(t)}{A_s}\right)^{2\beta}\right]^{\frac{1}{2\beta}}}.
\end{equation}

\subsubsection{HPA non-linearity and efficiency measurement}
\label{section_Measures_with_HPA_operation}
 {This section introduces several important metrics to characterize HPA's operating status  so that in the future analysis, we can build a relationship between the severity of HPA's non-linearity and the power harvesting performance as well as the  {trend of the optimal waveform's}. Firstly,  to show how HPA's non-linearity affects WPT's waveform preference, it would be beneficial to evaluate HPA's non-linearity numerically. An effective measurement is the output backoff (OBO) at HPA's operating point, which is defined as the ratio of HPA's saturation power {{$(P_{\text{s,HPA}})$}} to   { HPA's output signal power ($P_{\text{out,HPA}}$)} :
\begin{align}
\label{eq_OBO} 
\text{OBO}\overset{\bigtriangleup}{=}\frac{P_{\text{s,HPA}}}{ {P_{\text{out,HPA}}}}=\frac{A^2_s}{\mathcal{E}\left\lbrace{x}^{\text{HPA}^2}(t)\right\rbrace},
\end{align}
where ${x}^{\text{HPA}}(t)=|\widetilde{x}^{\text{HPA}}(t)|$ is the amplitude of the signal after HPA.  Smaller OBO indicates more severe HPA's non-linearity.

Besides the severity of non-linearity,  {power efficiency (PE)} is also an important metric to evaluate HPA's performance, defined as the ratio of  {HPA's output signal power ($P_{\text{out,HPA}}$) to HPA's DC power supply {$(P_{\text{DC,HPA}})$}}.  Here, we use a generalized PE approximation for all Class B HPAs (feature high PE) in \cite{ochiai2013analysis}:
\begin{align}
\label{eq_PE} 
\text{PE}\overset{\bigtriangleup}{=}\frac{ {P_{\text{out,HPA}}}}{P_{\text{DC,HPA}}}=\frac{\pi}{4}\frac{\mathcal{E}\left\lbrace    {x}^{\text{HPA}^2}(t)\right\rbrace}{A_s\mathcal{E}\left\lbrace    {x}^{\text{HPA}}(t)\right\rbrace}.
\end{align}

From Eq. \eqref{eq_PE}, the PE of HPA describes the efficiency of using HPA's DC power supply. Usually, higher input signal power drives more efficient use of HPA's DC power supply and produces higher output signal power, until HPA's saturation voltage is reached where the produced output power saturates.

One limitation of the PE metric in Eq. \eqref{eq_PE} is that it does not account for the power source from the input signal.  {When HPA approaches its saturation region, the output power is comparable to or even less than the input power of HPA, while the PE in Eq. \eqref{eq_PE} keeps increasing.}  In this context, a more effective metric of HPA efficiency is the added PE  {(APE)}, where the power of the input signal is also taken into consideration. APE is defined as the ratio of the produced signal power (the difference between  {HPA's output power and HPA's input power ($P_{\text{in,HPA}}$)}, i.e., $ P_{\text{ADE}}\overset{\bigtriangleup}{=}P_{\text{out,HPA}}-P_{\text{in,HPA}}$)  {to HPA's DC power supply}:
\begin{align}
\label{eq_APE} 
\text{APE}\overset{\bigtriangleup}{=}\frac{P_{\text{ADE}}}{P_{\text{DC,HPA}}}{=}\frac{\pi}{4}\frac{\mathcal{E}\left\lbrace    {x}^{\text{HPA}^2}(t)-{x}^{\text{in}^2}(t)\right\rbrace}{A_s\mathcal{E}\left\lbrace    {x}^{\text{HPA}}(t)\right\rbrace}.
\end{align}

APE gives us an insight  {into} HPA's PE considering all its power sources,  {including the input signal power and the DC power supply}. In contrast with the PE metric in Eq. \eqref{eq_PE} which prefers high input power, the APE in Eq. \eqref{eq_APE} might decrease or even become negative when the input signal power increases to HPA's highly non-linear operation regime, where the produced output power saturates regardless of the input signal power. In this condition, increasing the input signal power is not beneficial in terms of the whole system's  {overall} PTE, and APE would then become an important measurement metric. {The PE and APE metrics are introduced to give an insight into where the performance gain of the proposed waveforms comes from {when analyzing the simulations in section V.}}

\subsubsection{Analytical insights into HPA and BPF outputs}
\label{section_Analytical insights into HPA outputs}
In this section, we assume that the input waveform of HPA has the same activated frequencies as those of the transmit waveform, from $f_0$ to $f_{N-1}$ with frequency spacing $\Delta_f$. In this case, due to HPA's non-linearity, $\widetilde{x}^{\text{HPA}}(t)$ incurs distorted weights across the in-band frequencies and intermodulation leakage  {across} the out-of band frequencies at the multiples of $\Delta_f$. For the convenience of the following optimization formulation, we need to  {explicitly build the relationship between HPA's input waveform $\widetilde{x}^{\text{in}}(t)$, HPA's output waveform $\widetilde{x}^{\text{HPA}}(t)$ and the transmit waveform $\widetilde{x}^{\text{tr}}(t)$.}

Towards this, our main idea is to sample $\widetilde{x}^{\text{HPA}}(t)$ in Eq. \eqref{eq_HPA_model} and then conduct  {FT}.  {Starting} from Eq. \eqref{eq_HPA_model}, we have:
\begin{align}
\nonumber\widetilde{x}^{\text{HPA}}(t)&=f_{\text{HPA}}\left(\widetilde{x}^{\text{in}}_{B}(t)e^{j2\pi f_0 t}\right)=\frac{G\widetilde{x}^{\text{in}}_{B}(t)}{\left[1+\left(\frac{Gx^{\text{in}}_{B}(t)}{A_s}\right)^{2\beta}\right]^{\frac{1}{2\beta}}}e^{j2\pi f_0 t}\\\label{eq_HPA_model2}
&=\frac{G}{\left[1+\left(\frac{Gx^{\text{in}}_{B}(t)}{A_s}\right)^{2\beta}\right]^{\frac{1}{2\beta}}}\widetilde{x}^{\text{in}}(t)\overset{\Delta}{=}\mathcal{A}\left(x^{\text{in}}_{B}(t)\right)\widetilde{x}^{\text{in}}(t),
\end{align}
where $\widetilde{x}^{\text{in}}_{B}(t)=\widetilde{x}^{\text{in}}(t)e^{-2j\pi f_0t}$ is the baseband signal of $\widetilde{x}^{\text{in}}(t)$. Eq. \eqref{eq_HPA_model2} indicates that the spectrum of $\widetilde{x}^{\text{HPA}}(t)$ is the convolution between the spectrum of the baseband function $\mathcal{A}\left(x^{\text{in}}_{B}(t)\right)$ and the spectrum of $\widetilde{x}^{\text{in}}(t)$. The latter part is easily represented by $\left\lbrace\widetilde{w}^{\text{in}}_{n}\right\rbrace$ from Eq. \eqref{eq_input_signal_complex1}.

\textit{Remark 1:} We avoid sampling directly on $\mathcal{A}\left(x^{\text{in}}(t)\right)$ since a large number of samples would be required to achieve our desired frequency resolution $\Delta_f$ given its carrier frequency $f_0$,  {following the  Nyquist sampling theorem}. This will result in higher complexity or numerous equality constraints in the following optimization problem. On the other hand, sampling the baseband signal $\mathcal{A}\left(x^{\text{in}}_{B}(t)\right)$ is more practical.

To sample $\mathcal{A}\left(x^{\text{in}}_{B}(t)\right)$, we approximate its major frequency components are limited within $[0,~f^{\max}_{A}]$ with $f^{\max}_{A}=\kappa' N\Delta_f$. $\kappa'$ is the non-linear extending coefficient larger than 1. Then, based on the Nyquist sampling theorem, we set the sampling frequency at $2\kappa'N\Delta_f$. Further, to guarantee the frequency resolution of $\Delta_f$, we take $2\kappa'N$ samples, corresponding to a sampling interval of $t_s=T/2\kappa'N$ for a period of $T=1/\Delta_f$. Denote the $2\kappa'N$ samples of $ \mathcal{A}\left(x^{\text{in}}_{B}(t)\right)$ by $\left\lbrace A[k]\right\rbrace$  {for} $  k=0,~\cdots,~2\kappa'N-1$. From Eq. \eqref{eq_HPA_model2}, we have:
\begin{align}
\nonumber  \left\lbrace\widetilde{w}^{\text{HPA}}_{n}\right\rbrace=&\mathcal{F} \left\lbrace\widetilde{x}^{\text{HPA}}(t)\right\rbrace  |_{f=n\Delta_f+f_0}\\\nonumber
=&\mathcal{F}\left\lbrace \mathcal{A}\left(x^{\text{in}}_{B}(t)\right)\widetilde{x}^{\text{in}}(t)\right\rbrace|_{f=n\Delta_f+f_0}\\\label{eq_sspa_inverse_freq_1}
=&\frac{1}{2\kappa'N}\text{DFT}\left\lbrace A[k]\right\rbrace  \circledast_{2\kappa'N} \left\lbrace\widetilde{w}^{\text{in}}_{n}\right\rbrace,
\end{align}
with
\begin{equation}
\label{eq_A_k}
A[k]=\frac{G}{[1+(\frac{Gx^{\text{in}}_{B}(t_k)}{A_s})^{2\beta}]^{\frac{1}{2\beta}}},
\end{equation}
 {where $A[k]=  \mathcal{A}\left(x^{\text{in}}_{B}(t_k)\right)$ with $t_k=kt_s$. $n=0,~\cdots,~2\kappa'N-1$ for $\left\lbrace\widetilde{w}^{\text{HPA}}_{n}\right\rbrace$}.

From Eq. \eqref{eq_sspa_inverse_freq_1}, we get the real part of $\widetilde{w}^{\text{HPA}}_{n}$ as:
\begin{align}
\nonumber\overline{w}^{\text{HPA}}_{n}=&\frac{1}{2\kappa'N}\mathfrak{R} \left\lbrace\sum_{n_1=0}^{N-1} \widetilde{w}^{\text{in}}_{n_1}\sum_{k=0}^{2\kappa'N-1} A[k] e^{\frac{-j2\pi k(n-n_1)_{2\kappa'N}}{2\kappa'N}}\right\rbrace  \\\nonumber
=&\frac{1}{2\kappa'N}\sum_{n_1=0}^{N-1} \overline{w}^{\text{in}}_{n_1}\sum_{k=0}^{2\kappa'N-1} A[k] \cos\left(\frac{2\pi k(n-n_1)_{2\kappa'N}}{2\kappa'N}\right)\\\label{eq_sspa_inverse_freq2_real}
&-\frac{1}{2\kappa'N}\sum_{n_1=0}^{N-1} \widehat{w}^{\text{in}}_{n_1}\sum_{k=0}^{2\kappa'N-1} A[k] \sin\left(\frac{2\pi k(n-n_1)_{2\kappa'N}}{2\kappa'N}\right).
\end{align}

Similarly, for the imaginary part, we have:
\begin{align}
\nonumber\widehat{w}^{\text{HPA}}_{n}=&\frac{1}{{2\kappa'N}}\mathfrak{I} \left\lbrace\sum_{n_1=0}^{N-1} \widetilde{w}^{\text{in}}_{n_1}\sum_{k=0}^{2\kappa'N-1} A[k] e^{\frac{-j2\pi k(n-n_1)_{2\kappa'N}}{2\kappa'N}}\right\rbrace  \\\nonumber
=&\frac{1}{{2\kappa'N}}\sum_{n_1=0}^{N-1} \overline{w}^{\text{in}}_{n_1}\sum_{k=0}^{2\kappa'N-1}A[k] \sin\left(\frac{2\pi k(n-n_1)_{2\kappa'N}}{2\kappa'N}\right)\\\label{eq_sspa_inverse_freq2_imag}
&+\frac{1}{{2\kappa'N}}\sum_{n_1=0}^{N-1} \widehat{w}^{\text{in}}_{n_1}\sum_{k=0}^{2\kappa'N-1} A[k] \cos\left(\frac{2\pi k(n-n_1)_{2\kappa'N}}{2\kappa'N}\right).
\end{align}

 Assume that $\widetilde{x}^{\text{HPA}}(t)$ is then transmitted into an ideal BPF with its passband $[f_0,~f_0+(N-1)\Delta_f]$, and becomes the transmitted RF signal $\widetilde{x}^{\text{tr}}(t)$. {The transfer characteristics of this ideal  {BPF} is given as}:
\begin{align}
\label{eq_transmit_signal_complex}
\widetilde{w}^{\text{tr}}_{n}=\begin{cases}\widetilde{w}^{\text{HPA}}_{n}, &{n\in[0,...,N-1],}\\ 0,&\text{otherwise.}\end{cases}
\end{align}

 \subsection{Rectenna Model}
\begin{figure}
\centering
\includegraphics[scale=0.65]{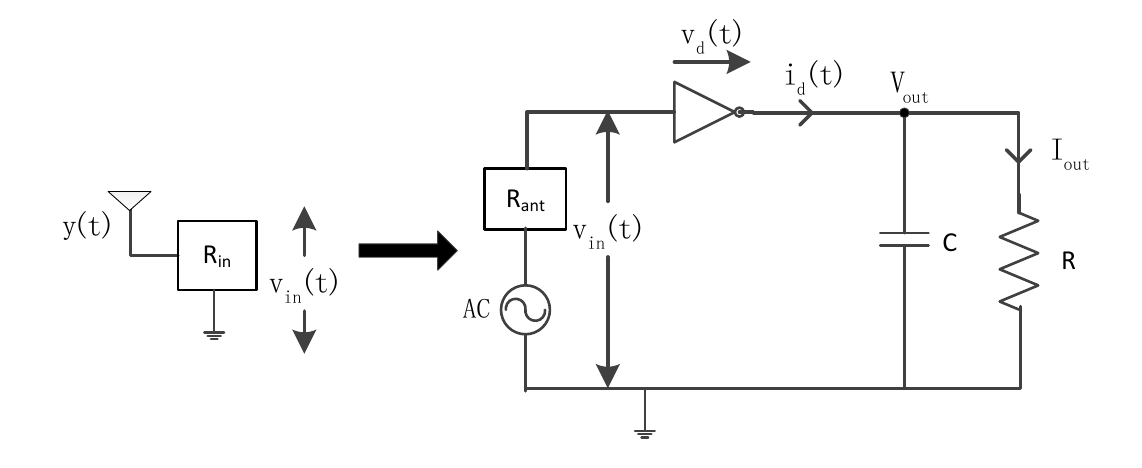}
\caption{The rectenna circuit.}
\label{rectenna}
\end{figure}
At the receiver, the wireless signal $\overline{y}(t)$ is picked up and is converted into DC via a rectenna.  We model the non-linear rectenna based on \cite{Clerckx2016Waveform} whose equivalent circuit is depicted in Fig. \ref{rectenna}. In this circuit, the receiving antenna, after picking up the RF signals, functions as a voltage source with average power $\mathcal{E}\left\lbrace\overline{y}(t)^2\right\rbrace  $ and an inner impedance $R_\text{ant}$. Denoting the source voltage as $v_s(t)$, we have $v_s(t)=2\overline{y}(t)\sqrt{R_\text{ant}}$.  {Assume that the rectenna circuit's input impedance $R_\text{in}$ matches with $R_\text{ant}$. Hence, the input voltage fed into the rectenna is $v_{\text{in}}(t)=v_s(t)/2=\overline{y}(t)\sqrt{R_\text{ant}}$, as shown on the right hand side of Fig. \ref{rectenna}.}

In Fig. \ref{rectenna}, the rectenna is  composed of a diode and a low pass filter (LPF)\footnote{We adopt the simplest resistor-capacitor LPF circuit at the EH. The LPF is assumed to be able to remove all the high-frequency harmonic components of the multi-sine signal after the non-linear diode at the rectenna.}.  {The transfer characteristic of the diode is written as $i_d(t)=i_s\exp\left\lbrace    v_d(t)/(nv_t)-1\right\rbrace $ with $v_d(t)=v_{\text{in}}(t)-V_{\text{out}}$  {($ V_{\text{out}}$ is the output DC voltage of the rectenna)} assuming a small signal model. To simplify the characteristic function in the optimization, the current $i_d(t)$ is  approximated by Taylor expansion at point $-V_{\text{out}}$. This gives $i_d(t)=\sum_{i=0}^{\infty}k_i v^i_{\text{in}}(t)=\sum_{i=0}^{\infty}k'_i R^{\frac{i}{2}}_{\text{ant}} \overline{y}^i(t)$}, where $k'_i$ is the $i^{\text{th}}$ order Taylor coefficient.  {Then, $i_d(t)$ passes through an LPF where its high-frequency components are filtered out before being fed into the load R. Assume that the LPF filters out all the high-frequency harmonic components in $i_d(t)$ ideally. Since the expectation of high-frequency components over time is $0$, at the load R, we have $I_{\text{out}}=\mathcal{E}\left\lbrace i_d(t)\right\rbrace=\sum_{i=0\text{, even}}^{i}k'_i R^i_{\text{ant}} \mathcal{E}\left\lbrace\overline{y}^i(t)\right\rbrace$ where the second equation comes from \cite{Clerckx2016Waveform}}. To maximize the generated power at the load R is equivalent to maximize the current at the load R, i.e., $I_{\text{out}}$, which, from \cite{Clerckx2016Waveform}, is in proportion to the scaling term  {(as a function of the transmit waveform $ \left\lbrace\overline{w}^{\text{tr}}_{n}\right\rbrace,~  \left\lbrace\widehat{w}^{\text{tr}}_{n}\right\rbrace$)}:
\begin{subequations}
\begin{align}
\nonumber\label{eq_scaling_term0}
& {z_{\text{DC}}\left( \left\lbrace\overline{w}^{\text{tr}}_{n}\right\rbrace,~  \left\lbrace\widehat{w}^{\text{tr}}_{n}\right\rbrace\right)}\\
\overset{\Delta}{=}&k_2R_{\text{ant}}\mathcal{E}\left\lbrace\overline{y}(t)^2\right\rbrace  +k_4R_{\text{ant}}^2\mathcal{E}\left\lbrace\overline{y}(t)^4\right\rbrace  \\
\label{eq_SSPA_poly_x_tr}
\nonumber=&\frac{k_2R_{\text{ant}}}{2}\sum_{n=0}^{N-1}|\widetilde{w}^{\text{tr}}_{n}\widetilde{h}_{n}|^2+\frac{3k_4R_{\text{ant}}^2}{8}(\sum_{\tiny{\begin{array}{c} n_0,n_1,n_2,n_3\\n_0+n_1=n_2+n_3\end{array}}} \widetilde{h}_{n_0}\widetilde{w}^{\text{tr}}_{n_0}\times\\
&\:\:\:\:\:\:\:\:\:\:\:\:\:\:\:\:\:\:\:\:\widetilde{h}_{n_1}\widetilde{w}^{\text{tr}}_{n_1}\widetilde{h}^*_{n_2}\widetilde{w}^{\text{tr}^*}_{n_2}\widetilde{h}^*_{n_3}\widetilde{w}^{\text{tr}^*}_{n_3}),
\end{align}
\end{subequations}
 {with $k_2=0.0034$ and $k_4=0.3829$}. Eq. \eqref{eq_scaling_term0} only reserve{s} the second\footnote{If only reserving the second order Taylor expansion term, the generated power at the load R would be a linear function of the received RF signals,  in which case the preferred waveform for the EH shifts from multi-carrier transmission to single-carrier transmission \cite{Clerckx2016Waveform}. Considering HPA's preference for low-PAPR waveforms, there would be no trade-off  {on waveform design} between the linear EH and the non-linear HPA, and the optimal waveform becomes single-carrier transmission regardless of HPA's non-linearity.} and the fourth order terms of the Taylor approximation, which have been able to capture the non-linear properties of the rectenna.

Based on the scaling term $z_{\text{DC}}$, the scaling term of the end-to-end PTE of the system is defined as:
\begin{equation}
\label{eq_E2E_PTE}
\text{PTE}=\frac{z^2_{\text{DC}}R}{ {P_{\text{in,HPA}}}+P_{\text{DC,HPA}}},
\end{equation}
where $ {P_{\text{in,HPA}}}+P_{\text{DC,HPA}}$ refers to all the  power supply at the transmitter\footnote{We use   $z_{\text{DC}}$ to represent the harvested power in the paper because it is directly in  {proportion} to the generated DC current at the load and thus is in  {proportion} to the harvested power. Similarly, the PTE defined in Eq. \eqref{eq_E2E_PTE} is in  {proportion} to the real end-to-end PTE and will represent the end-to-end PTE in the paper.  The end-to-end PTE is put forward to highlight an appropriate operating region of the WPT system, where both a satisfying harvested energy and an efficient power transfer efficiency are achieved.}.

\section{Optimization Solutions}
In this section, we first formulate a generalized optimization problem with respect to the input waveform of HPA  { to maximize the harvested power}, subjective to the input\footnote{The input power constraint at HPA avoids poor amplifier PE which occurs when the power of HPA's input signal gets close to or exceeds HPA's saturation power.} and transmit power constraints\footnote{The transmit power constraint at antennas limits the RF exposure to human beings over propagation.}.  Then, for tractability, we specify and solve the generalized optimization model based on two different assumptions on the frequencies of the input waveform at HPA:

1) Model I: the input waveform at HPA has  {the same multi-carrier frequencies as the transmit waveform within the transmit pass band}.

2) Model II: the input waveform at HPA uses the frequencies out of the transmit pass band to adaptively design the input waveform that can pass the BPF after HPA in a lossless way.
\label{section_optimization}
\subsection{Problem Formulation}
We first formulate a general form of the optimization problem to maximize the harvested power in WPT:
\begin{maxi!}
        { {\left\lbrace  \overline{w}^{\text{in}}_{n}\right\rbrace,~  \left\lbrace\widehat{w}^{\text{in}}_{n}\right\rbrace}}{ z_{\text{DC}}\left( \left\lbrace\overline{w}^{\text{tr}}_{n}\right\rbrace,~ \left\lbrace\widehat{w}^{\text{tr}}_{n}\right\rbrace\right),}{\label{eq_optimization_P1}}{\label{eq_optimization_P1_1}}
        \addConstraint{ \frac{1}{2} \sum_{n\in \kappa^{\text{in}}} |\overline{w}^{\text{in}}_{n}|^2 +|\widehat{w}^{\text{in}}_{n}|^2\leq P^{\max}_{\text{in}}}\label{eq_optimization_P1_2}
        \addConstraint{ \frac{1}{2} \sum_{n=0}^{N-1} |\overline{w}^{\text{tr}}_{n}|^2 + |\widehat{w}^{\text{tr}}_{n}|^2  \leq P^{\max}_{\text{tr}},}\label{eq_optimization_P1_3}
      \end{maxi!}
where $P^{\max}_{\text{in}}$ and $P^{\max}_{\text{tr}}$ are the input power constraint and the transmit power constraint respectively.

The main difficulty of solving problem \eqref{eq_optimization_P1} is that, the scaling term $z_{\text{DC}}$, as a non-linear function of $\left(\left\lbrace \overline{w}^{\text{tr}}_{n}\right\rbrace,~  \left\lbrace\widehat{w}^{\text{tr}}_{n}\right\rbrace\right)$  in Eq. \eqref{eq_SSPA_poly_x_tr}, involves extremely high-complexity if  {written} as a function of $\left( \left\lbrace\overline{w}^{\text{in}}_{n}\right\rbrace, ~ \left\lbrace\widehat{w}^{\text{in}}_{n}\right\rbrace\right)$ by adopting the waveform relationship modelled in section \ref {section_Measures_with_HPA_operation}. The same issue occurs for the transmit power constraint in Eq. \eqref{eq_optimization_P1_3}. Hence, two possible solutions we propose in the following are: 1) similarly to slack variables, we introduce $\left( \left\lbrace\overline{w}^{\text{tr}}_{n}\right\rbrace,~ \left\lbrace\widehat{w}^{\text{tr}}_{n}\right\rbrace\right)$ as auxiliary variables, to avoid expressing the objective function (also the transmit power constraint) as a function of the input waveform, with the non-linear relationship between $\left( \left\lbrace\overline{w}^{\text{tr}}_{n}\right\rbrace,~ \left\lbrace\widehat{w}^{\text{tr}}_{n}\right\rbrace\right)$ and $\left( \left\lbrace\overline{w}^{\text{in}}_{n}\right\rbrace,~ \left\lbrace\widehat{w}^{\text{in}}_{n}\right\rbrace\right)$ represented by the non-linear equalities (Model I); 2) we represent the input power constraint in Eq. \eqref{eq_optimization_P1_2}  as a function of the transmit waveform $\left( \left\lbrace\overline{w}^{\text{tr}}_{n}\right\rbrace,~  \left\lbrace\widehat{w}^{\text{tr}}_{n}\right\rbrace\right)$, and make problem \eqref{eq_optimization_P1} a function of $\left( \left\lbrace\overline{w}^{\text{tr}}_{n}\right\rbrace,~ \left\lbrace\widehat{w}^{\text{tr}}_{n}\right\rbrace\right)$ instead (Model II). To achieve this, we assume that the frequencies of  {HPA's output, $\widetilde{x}^{\text{HPA}}(t)$, are concentrated within} the pass band of BPF, which makes BPF lossless and avoids characterizing the inverse function of the BPF. Consequently, we can write the input power constraint with respect to $\left( \left\lbrace\overline{w}^{\text{tr}}_{n}\right\rbrace,~  \left\lbrace\widehat{w}^{\text{tr}}_{n}\right\rbrace\right)$ explicitly. Correspondingly, to concentrate the frequencies of $\widetilde{x}^{\text{HPA}}(t)$ within BPF's pass band after the non-linear HPA,  {the corresponding HPA's input, $\widetilde{x}^{\text{HPA}}(t)$, will have  out-of band frequencies.} Details are given as follows.

\subsection{Model I: Input Waveform with the Same Frequencies as the Transmit Waveform}
As it has been mentioned, Model I solves the optimization problem where the input waveform at HPA assumes to have the same frequencies as the transmit waveform. In this context, for tractability, we collect both the input waveform of HPA $\left(\left\lbrace \overline{w}^{\text{in}}_{n}\right\rbrace,~ \left\lbrace\widehat{w}^{\text{in}}_{n}\right\rbrace\right)$ and the transmit waveform $\left(\left\lbrace \overline{w}^{\text{tr}}_{n}\right\rbrace,~ \left\lbrace\widehat{w}^{\text{tr}}_{n}\right\rbrace\right)$ into optimization variables so that the objective function and the transmit power constraint are in a much simpler form. The relationship between the transmit waveform and the input waveform is derived in Eq. \eqref{eq_sspa_inverse_freq2_real} and Eq. \eqref{eq_sspa_inverse_freq2_imag}. Intuitively, the auxiliary variables $\left(\left\lbrace \overline{w}^{\text{tr}}_{n}\right\rbrace,~ \left\lbrace\widehat{w}^{\text{tr}}_{n}\right\rbrace\right)$ enable us to handle the original high complexity in  {Eq. \eqref{eq_optimization_P1_1}} and Eq. \eqref{eq_optimization_P1_3} separately.  The corresponding optimization problem is written as:
  \begin{maxi!}
        {\substack{\tiny{\begin{matrix} \left\lbrace\overline{w}^{\text{in}}_{n}\right\rbrace,~ \left\lbrace\widehat{w}^{\text{in}}_{n}\right\rbrace  \\ \left\lbrace\overline{w}^{\text{tr}}_{n}\right\rbrace,~ \left\lbrace\widehat{w}^{\text{tr}}_{n}\right\rbrace  \end{matrix}}}}{z_{\text{DC}}\left( \left\lbrace\overline{w}^{\text{tr}}_{n}\right\rbrace,~ \left\lbrace\widehat{w}^{\text{tr}}_{n}\right\rbrace\right),}{\label{eq_optimization_P2}}{\label{eq_optimization_P2_1}}
        \addConstraint{\frac{1}{2} \sum_{n=0}^{N-1} {\overline{w}^{\text{in}^2}_{n}}+\widehat{w}^{\text{in}^2}_{n} \leq P^{\max}_{\text{in}}}\label{eq_optimization_P2_3}
       \addConstraint{\frac{1}{2} \sum_{n=0}^{N-1} {\overline{w}^{\text{tr}^2}_{n}}+\widehat{w}^{\text{tr}^2}_{n} \leq P^{\max}_{\text{tr}}}\label{eq_optimization_P2_4}
        \addConstraint{\overline{w}^{\text{HPA}}_{n}\text{ in Eq. \eqref{eq_sspa_inverse_freq2_real}}}\label{eq_optimization_P2_6}
       \addConstraint{\widehat{w}^{\text{HPA}}_{n}\text{ in Eq. \eqref{eq_sspa_inverse_freq2_imag}},}\label{eq_optimization_P2_7}
      \end{maxi!}
where the constraints in Eq.\eqref{eq_optimization_P2_6} and Eq.\eqref{eq_optimization_P2_7} feature high non-linearity.

Problem \eqref{eq_optimization_P2} maximizes a convex objective (Appendix A), which can be efficiently solved by SCP. In SCP, the objective term is linearly approximated by its first-order Taylor expansion at a fixed operating point, forming a new tractable optimization problem whose optimal solution is used as a new operating point of the next iteration. The procedure is repeated until two successive solutions are close enough and can be viewed as the solution of problem \eqref{eq_optimization_P2}. Assume $\left( \left\lbrace\overline{w}^{\text{in},(l-1)}_{n}\right\rbrace,~ \left\lbrace\widehat{w}^{\text{in},(l-1)}_{n}\right\rbrace,~ \left\lbrace\overline{w}^{\text{tr},(l-1)}_{n}\right\rbrace,~ \left\lbrace\widehat{w}^{\text{tr},(l-1)}_{n}\right\rbrace\right)$ are the values of the operating point at the beginning of the $l^{\text{th}}$ iteration. Then, $z_{\text{DC}}\left(\left\lbrace\overline{w}^{\text{tr}}_{n}\right\rbrace,~ \left\lbrace\widehat{w}^{\text{tr}}_{n}\right\rbrace\right)$ at the $l^{\text{th}}$ iteration is linearly approximated as:
\begin{align}
\label{eq_first_order_Taylor}
z_{\text{DC}}^{(l)}\left( \left\lbrace\overline{w}^{\text{tr}}_{n}\right\rbrace,~ \left\lbrace\widehat{w}^{\text{tr}}_{n}\right\rbrace\right)=\sum_{n=0}^{N-1} \overline{\alpha}^{(l)}_{n}\overline{w}^{\text{tr}}_{n}+\widehat{\alpha}^{(l)}_{n}\widehat{w}^{\text{tr}}_{n},
\end{align}
where $\left(\left\lbrace\overline{\alpha}^{(l)}_{n}\right\rbrace,~\left\lbrace\widehat{\alpha}^{(l)}_{n}\right\rbrace  \right)$ are the first-order Taylor coefficients at the $l^{\text{th}}$ iteration, shown in Eq.\eqref{eq_alpha_1} and \eqref{eq_alpha_2}.

\begin{figure*}[t]
\begin{align}
\nonumber\overline{\alpha}^{(l)}_{n}=&{k_2 R_{\text{ant}}}\overline{w}^{\text{tr},(l-1)}_{n}|\widetilde{h}_{n}|^2+3/8k_4R_{\text{ant}}\left[\overline{w}^{{\text{tr},(l-1)}}_{n}|\widetilde{w}^{{\text{tr},(l-1)}}_{n}|^2|\widetilde{h}_{n}|^4+\sum_{n_1}8\overline{w}^{{\text{tr},(l-1)}}_{n}|\widetilde{w}^{{\text{tr},(l-1)}}_{n_1}|^2|\widetilde{h}_{n}|^2|\widetilde{h}_{n_1}|^2+\right.\\\label{eq_alpha_1} &\left.\sum_{\tiny{\begin{array}{c} n_2+n_3=2n\\n_2\neq n_3\end{array}}}4\mathfrak{R}\left\lbrace\widetilde{w}^{\text{tr},(l-1)}_{n}\widetilde{w}^{{\text{tr},(l-1)}^*}_{n_2}\widetilde{w}^{{\text{tr},(l-1)}^*}_{n_3}\widetilde{h}^2_{n}\widetilde{h}^*_{n_2}\widetilde{h}^*_{n_3}\right\rbrace+\sum_{\tiny{\begin{array}{c} n=-n_1+n_2+n_3\\n\neq n_1\neq n_2\neq n_3\end{array}}}4\mathfrak{R}\left\lbrace\widetilde{w}^{{\text{tr},(l-1)}}_{n_1}\widetilde{w}^{{\text{tr},(l-1})^*}_{n_2}\widetilde{w}^{{\text{tr},(l-1)}^*}_{n_3}\widetilde{h}_{n}\widetilde{h}_{n_1}\widetilde{h}^*_{n_2}\widetilde{h}^*_{n_3}\right\rbrace\right],
\end{align}
\begin{align}
\nonumber\widehat{\alpha}^{(l)}_{n}=&{k_2 R_{\text{ant}}}\widehat{w}^{\text{tr},(l-1)}_{n}|\widetilde{h}_{n}|^2+{3}/{8}k_4R_{\text{ant}}\left[4\widehat{w}^{{\text{tr},(l-1)}}_{n}|\widetilde{w}^{{\text{tr},(l-1)}}_{n}|^2|\widetilde{h}_{n}|^4+\sum_{n_1}8\widehat{w}^{{\text{tr},(l-1)}}_{n}|\widetilde{w}^{{\text{tr},(l-1)}}_{n_1}|^2|\widetilde{h}_{n}|^2|\widetilde{h}_{n_1}|^2-\right.\\\label{eq_alpha_2}
&\left.\sum_{\tiny{\begin{array}{c} n_2+n_3=2n\\n_2\neq n_3\end{array}}}4\mathfrak{I}\left\lbrace\widetilde{w}^{\text{tr},(l-1)}_{n}\widetilde{w}^{{\text{tr},(l-1)}^*}_{n_2}\widetilde{w}^{{\text{tr},(l-1)}^*}_{n_3}\widetilde{h}^2_{n}\widetilde{h}^*_{n_2}\widetilde{h}^*_{n_3}\right\rbrace-\sum_{\tiny{\begin{array}{c} n=-n_1+n_2+n_3\\n\neq n_1\neq n_2\neq n_3\end{array}}}4\mathfrak{I}\left\lbrace\widetilde{w}^{{\text{tr},(l-1)}}_{n_1}\widetilde{w}^{{\text{tr},(l-1})^*}_{n_2}\widetilde{w}^{{\text{tr},(l-1)}^*}_{n_3}\widetilde{h}_{n}\widetilde{h}_{n_1}\widetilde{h}^*_{n_2}\widetilde{h}^*_{n_3}\right\rbrace\right].
\end{align}
\noindent\makebox[\linewidth]{\rule{\paperwidth}{0.4pt}}
\end{figure*}

Correspondingly,   problem \eqref{eq_optimization_P2} can be transformed into:
  \begin{maxi!}
        {\substack{\tiny{\begin{matrix} \left\lbrace\overline{w}^{\text{in}}_{n}\right\rbrace,~ \left\lbrace\widehat{w}^{\text{in}}_{n}\right\rbrace  \\\left\lbrace\overline{w}^{\text{tr}}_{n}\right\rbrace,~\left\lbrace\widehat{w}^{\text{tr}}_{n}\right\rbrace  \end{matrix}}}}{z_{\text{DC}}^{(l)}(\left\lbrace\overline{w}^{\text{tr}}_{n}\right\rbrace,~\left\lbrace\widehat{w}^{\text{tr}}_{n}\right\rbrace),}{\label{eq_optimization_P5}}{\label{eq_optimization_P5_1}}
        \addConstraint{g_i^\text{I}( \left\lbrace\overline{w}^{\text{in/tr}}_{n}\right\rbrace,~\left\lbrace\widehat{w}^{\text{in/tr}}_{n}\right\rbrace)\leq 0, ~i=\left\lbrace 1,~2\right\rbrace}\label{eq_optimization_P5_3}
        \addConstraint{g_i^\text{E}( \left\lbrace\overline{w}^{\text{in/tr}}_{n}\right\rbrace,~ \left\lbrace\widehat{w}^{\text{in/tr}}_{n}\right\rbrace)= 0,~i=\left\lbrace    1,~\cdots,~2N\right\rbrace,}\label{eq_optimization_P5_4}
      \end{maxi!}
where $g_{1/2}^\text{I}( \left\lbrace\overline{w}^{\text{in/tr}}_{n}\right\rbrace,~\left\lbrace\widehat{w}^{\text{in/tr}}_{n}\right\rbrace)$ refers to the inequality constraint in Eq. \eqref{eq_optimization_P2_3} and Eq. \eqref{eq_optimization_P2_4} respectively. $g_{i}^\text{E}( \left\lbrace\overline{w}^{\text{in/tr}}_{n}\right\rbrace,~\left\lbrace\widehat{w}^{\text{in/tr}}_{n}\right\rbrace)$ refers to all the equality constraints in Eq. \eqref{eq_optimization_P2_6} and Eq. \eqref{eq_optimization_P2_7}.

Problem \eqref{eq_optimization_P5} can be solved by SQP in a highly efficient way. SQP is extended from solving the KKT conditions of an arbitrary non-linear program by using the Newton-Raphson (NR) method (named by KKT-NR) \cite{nocedal1999numerical}. To summarize, in KKT-NR, the NR method is used to solve the KKT functions iteratively with a closed-form solution over each iteration, and shows a fast-convergence property. In this context, SQP constructs a linearly constrained quadratic problem (LCQP) which has the same solution and which approximates the original program at the initialized point at each iteration. In this way, SQP achieves the convergence property of the KKT-NR's method. Hence in SQP, the original non-linear programming is solved by iterative approximations, where in each iteration, only a low-complex LCQP problem needs to be solved. Specifically, in each iteration, we get the direction of the optimized variables from the sub-program:
\begin{maxi!}
        {\substack{\bigtriangleup\mathbf{w}}}{\bigtriangledown ^T z_{\text{DC}}^{(l)}(\mathbf{w}_k)\bigtriangleup\mathbf{ w}+\frac{\bigtriangleup\mathbf{ w}^T H^{(l)}(\mathbf{w}_k)\mathbf{\bigtriangleup w}}{2},}{\label{eq_optimization_P8}}{\label{eq_optimization_P8_1}}
        \addConstraint{\bigtriangledown^T g_i^\text{I}(\mathbf{w}_k)\bigtriangleup\mathbf{w}+ g_i^\text{I}(\mathbf{w}_k)\leq 0,~i=\left\lbrace    1,~2\right\rbrace}\label{eq_optimization_P8_2}
          \addConstraint{\bigtriangledown^T g_i^\text{E}(\mathbf{w}_k)\bigtriangleup\mathbf{w}+ g_i^\text{E}(\mathbf{w}_k)=0,~i=\left\lbrace    1,~\cdots,~2N\right\rbrace,}\label{eq_optimization_P8_3}
      \end{maxi!}
where $\mathbf{w}=[{\overline{\mathbf{w}}}^{\text{in}^T},~{\widehat{\mathbf{w}}}^{\text{in}^T},~{\overline{\mathbf{w}}}^{\text{tr}^T},~{\widehat{\mathbf{w}}}^{\text{tr}^T}]^T$ with ${\overline{\mathbf{w}}}^{\text{in/tr}^T}=[\overline{w}^{\text{in/tr}}_{0},~\cdots,~\overline{w}^{\text{in/tr}}_{N-1}]^T$  and ${\widehat{\mathbf{w}}}^{\text{in/tr}^T}=[\widehat{w}^{\text{in/tr}}_{0},~\cdots,~\widehat{w}^{\text{in/tr}}_{N-1}]^T$. $H^{(l)}(\mathbf{w}_k)$ is the Hermitian matrix of the Lagrangian function in problem \eqref{eq_optimization_P5} at the initialized point $\mathbf{w_k}$:
\begin{align}
\label{eq_Hermain_Langrangian}
&H^{(l)}(\mathbf{w_k})_{j_1,~j_2}=\frac{\partial^2 L_k^{(l)}(\mathbf{w})}{\partial w_{j_1} \partial w_{j_2}}|_{\mathbf{w}=\mathbf{w_k}},\\
&L_k^{(l)}(\mathbf{w})= z_{\text{DC}}^{(l)}(\mathbf{w})+\sum_{i=1}^{2N} u^k_i g_i^\text{E}(\mathbf{w})+ \sum_{i=1}^{2} v^k_i g_i^\text{I}(\mathbf{w}),
\end{align}
where $u^k_i~(i=[1,~\cdots,~2N])$ and $v^k_i~(i=[1,~2])$ are the initializations of the Lagrangian multipliers at each iteration, which are obtained from the Lagrangian multipliers of the LCQP sub-program in the previous iteration.

Problem \eqref{eq_optimization_P8} is an easily handled LCQP problem. When $\bigtriangleup\mathbf{w}$ converges to $\mathbf{0}$, the sub-program in \eqref{eq_optimization_P8} converges to the original program as is easily seen \cite{boggs1995sequential}.

Since SQP derives from the NR method to solve KKT conditions, it also preserves the features of the KKT-NR method, i.e., it is not a feasible-point (points that satisfy the constraints) method which means that the results from its sub-programs  {in Eq. \eqref{eq_optimization_P8}} might not satisfy all the constraints in the original program. This is beneficial in the sense that the initialized point can be picked more randomly. However, it also gives probability that one of the sub-program gives a solution far from being feasible and requires more iterations to come back (In this condition, terminating the iteration in advance might result in worse performance). This drawback makes  the results susceptible to the initializations and the functions' derivative features, especially in large-scale problems or in problems with numerous constraints. Hence, when solving problem \eqref{eq_optimization_P2}, we introduce auxiliary variables  {and handle the corresponding objective function iteratively by SCP}. In this way, we construct simpler sub-programs to be solved by SQP and get more sophisticated initialization points in each iteration. The structure is named by SCP-SQP as in Algorithm \ref{SCP-SQP}.

\begin{algorithm}[t]
$\textbf{Input}$: $( \left\lbrace\overline{w}^{\text{in}}_{n}\right\rbrace,~ \left\lbrace\widehat{w}^{\text{in}}_{n}\right\rbrace,~\left\lbrace\overline{w}^{\text{tr}}_{n}\right\rbrace,~\left\lbrace\widehat{w}^{\text{tr}}_{n}\right\rbrace)^{(0)},~\epsilon_0>0,~l\leftarrow 1$;\\
 $\textbf{Output}$: $( \left\lbrace\overline{w}^{\text{in}}_{n}\right\rbrace,~\left\lbrace\widehat{w}^{\text{in}}_{n}\right\rbrace,~\left\lbrace\overline{w}^{\text{tr}}_{n}\right\rbrace,~\left\lbrace\widehat{w}^{\text{tr}}_{n}\right\rbrace)^{\star}$;\\
  $\textbf{Repeat}$: \\
1:Compute $ ( \left\lbrace\overline{\alpha}_{n}\right\rbrace,~\left\lbrace\widehat{\alpha}_{n}\right\rbrace )^{(l)} $  using Eq. \eqref{eq_alpha_1} and Eq. \eqref{eq_alpha_2} at the operating point $\left( \left\lbrace\overline{w}^{\text{tr}}_{n}\right\rbrace,~\left\lbrace\widehat{w}^{\text{tr}}_{n}\right\rbrace\right)^{(l-1)}$;\\
2:Compute $( \left\lbrace\overline{w}^{\text{in}}_{n}\right\rbrace,~ \left\lbrace\widehat{w}^{\text{in}}_{n}\right\rbrace,~\left\lbrace\overline{w}^{\text{tr}}_{n}\right\rbrace,~\left\lbrace\widehat{w}^{\text{tr}}_{n}\right\rbrace)^{(l)}$ in problem \eqref{eq_optimization_P5} using SQP in Algorithm \ref{algorithm_sqp};\\
3:Update $( \left\lbrace\overline{w}^{\text{in}}_{n}\right\rbrace,~ \left\lbrace\widehat{w}^{\text{in}}_{n}\right\rbrace,~\left\lbrace\overline{w}^{\text{tr}}_{n}\right\rbrace,~\left\lbrace\widehat{w}^{\text{tr}}_{n}\right\rbrace)^{\star}$ as the current solution;\\
4:Quit if
$|z_{\text{DC}}( \left\lbrace\overline{w}^{\text{tr}}_{n}\right\rbrace,~ \left\lbrace\widehat{w}^{\text{tr}}_{n}\right\rbrace)^{(l)}-z_{\text{DC}}\left( \left\lbrace\overline{w}^{\text{tr}}_{n}\right\rbrace,~\left\lbrace\widehat{w}^{\text{tr}}_{n}\right\rbrace\right)^{(l-1)}|< \epsilon_0$;\\
5:$l\leftarrow l+1$;
\caption{SCP-SQP}
\label{SCP-SQP}
\end{algorithm}
\begin{algorithm}[t]
 $\textbf{Input}$: $k=0,~\mathbf{w}^{(k)}\leftarrow\mathbf{w}^{(l-1)},~\{u^{k}_i\},\{v^{k}_i\},~\epsilon_S>0,~\alpha$;\\
 $\textbf{Output}$: $\mathbf{w}^{(l)}$; \\
 $\textbf{Repeat}$: \\
 1:Compute $H^{(l)}(\mathbf{w}^{(k)})$ (Eq. \eqref{eq_Hermain_Langrangian}), $\bigtriangledown z_{\text{DC}}^{(l)}(\mathbf{w}^{(k)})$,  $g_i^\text{I}(\mathbf{w}^{(k)})$, $\bigtriangledown g_i^\text{I}(\mathbf{w}^{(k)})$, $g_i^\text{E}(\mathbf{w}^{(k)})$, and $\bigtriangledown g_i^\text{E}(\mathbf{w}^{(k)})$;\\
2:Compute $\bigtriangleup \mathbf{w}$, $\{u^{k+1}_i\}$ and $\{v^{k+1}_i\}$ by solving the LCQP in \eqref{eq_optimization_P8};\\
3:$\mathbf{w}^{(k+1)}\leftarrow \mathbf{w}^{(k)}+\bigtriangleup \mathbf{w}$;\\
4:Quit if $\| \bigtriangleup \mathbf{w}\|< \epsilon_S, ~\mathbf{w}^{(l)}\leftarrow \mathbf{w}^{(k+1)}$;\\
5:$k\leftarrow k+1$;\;
\caption{SQP}
\label{algorithm_sqp}
\end{algorithm}

\textit{Remark 2:} The solution in Model I can also be applied to other HPA models or to more sophisticated architectures in WPT, such as when considering clipping DACs at the transmitter. First, using DFT and sampling to develop the relationship between waveforms at different stages is easy to be extended to other architectures. Second, the proposed SCP-SQP algorithm is also robust to solving various non-linear programs and can be extended to multi-input multi-output or multi-user scenarios.

\subsection{Model II: Input Waveform Utilizing Out-of-Band Frequencies}
\label{section_opt1}
We now take advantage of the out-of band frequencies at HPA's input so that after HPA, the signal $x^{\text{HPA}}(t)$ can pass the BPF in a lossless way. Another benefit of this model is that under a lossless BPF scenario, the input power can be written as a function of the transmit waveform, the same as the objective function and the transmit power constraint. Hence, compared with Model I, Model II formulates the optimization problem with respect to the transmit waveform only, which releases the numerous non-linear equality constraints between the input waveform and the transmit waveform. Specifically, we form the optimization problem as:
      \begin{maxi!}
       {\substack{\left\lbrace \overline{w}^{\text{tr}}_{n}\right\rbrace,~\left\lbrace\widehat{w}^{\text{tr}}_{n}\right\rbrace}}{z_{\text{DC}}\left( \left\lbrace\overline{w}^{\text{tr}}_{n}\right\rbrace,~\left\lbrace\widehat{w}^{\text{tr}}_{n}\right\rbrace\right),}{\label{eq_optimization_P3}}{\label{eq_optimization_P3_1}}
        \addConstraint{\frac{1}{2TG^2}\int_{T}\left[{{x^{\text{tr}}(t)}^{-2\beta}-A_s^{-2\beta}}\right]^{-\frac{1}{\beta}}dt \leq P^{\max}_{\text{in}}}\label{eq_optimization_P3_3}
      \addConstraint{\frac{1}{2} \sum_{n=0}^{N-1} {\overline{w}^{\text{tr}^2}_{n}}+\widehat{w}^{\text{tr}^2}_{n} \leq P^{\max}_{\text{tr}},}\label{eq_optimization_P3_2}
     \end{maxi!}
where the constraint in Eq. \eqref{eq_optimization_P3_3} is shown to be the input power constraint in Appendix B. Eq. \eqref{eq_optimization_P3_3} can be proved convex in Appendix C.

Problem \eqref{eq_optimization_P3} maximizes a convex objective function, which can be solved by SCP. Similarly to the SCP-SQP method, at the $l^{\text{th}}$ iteration, problem \eqref{eq_optimization_P3} is approximated as:
\begin{maxi!}
        {\substack{ \left\lbrace\overline{w}^{\text{tr}}_{n}\right\rbrace,~\left\lbrace\widehat{w}^{\text{tr}}_{n}\right\rbrace}}{z_{\text{DC}}^{(l)}\left( \left\lbrace\overline{w}^{\text{tr}}_{n}\right\rbrace,~\left\lbrace\widehat{w}^{\text{tr}}_{n}\right\rbrace\right),}{\label{eq_optimization_P4}}{\label{eq_optimization_P4_1}}
        \addConstraint{\text{Eq}. \eqref{eq_optimization_P3_2},\quad \text{Eq}. \eqref{eq_optimization_P3_3}.}{\label{eq_optimization_P4_2}}
      \end{maxi!}

Problem \eqref{eq_optimization_P4} is a convex problem with a linear objective function and two non-linear convex inequality constraints. However, problem \eqref{eq_optimization_P4} involves a highly non-linear inequality constraint in Eq. \eqref{eq_optimization_P3_2}, which can be efficiently solved by the IP method. The IP method uses the Barrier's method to bring the inequality constraints into the objective function equivalently and then solves the resultant program by using Gradient Descend (GD) method.

Specifically, the non-linear constraints in Eq. \eqref{eq_optimization_P4_2}, according to the Barrier's method, are omitted by reformulating problem \eqref{eq_optimization_P4} into:
\begin{align}
\label{eq_optimization_P4_l}
\min_{\left\lbrace\overline{w}^{\text{tr}}_{n}\right\rbrace,~\left\lbrace\widehat{w}^{\text{tr}}_{n}\right\rbrace} \quad  &-z_{\text{DC}}^{(l)}(\left\lbrace \overline{w}^{\text{tr}}_{n}\right\rbrace,~\left\lbrace\widehat{w}^{\text{tr}}_{n}\right\rbrace) +\sum_{i=1}^{2}I_-(f_{c,i}(\left\lbrace\overline{w}^{\text{tr}}_{n}\right\rbrace,~\left\lbrace\widehat{w}^{\text{tr}}_{n}\right\rbrace)),
\end{align}
where
\begin{align}
\label{eq_interpratation}
I_-(x)=&\lbrace\begin{matrix}
0,\:\:\:\:&x\leq 0,\\
\infty,\:\:\:\:&x> 0,
\end{matrix}\\\label{eq_interpratation1}
f_{c,1}(\left\lbrace\overline{w}^{\text{tr}}_{n}\right\rbrace,~\left\lbrace\widehat{w}^{\text{tr}}_{n}\right\rbrace)=&\frac{\int_{T}\left[{{x^{\text{tr}}(t)}^{-2\beta}-A_s^{-2\beta}}\right]^{-\frac{1}{\beta}}dt}{2TG^2}- P^{\max}_{\text{in}},\\
 f_{c,2}(\left\lbrace\overline{w}^{\text{tr}}_{n}\right\rbrace,~\left\lbrace\widehat{w}^{\text{tr}}_{n}\right\rbrace)=&\frac{1}{2} \sum_{n=0}^{N-1} {\overline{w}^{\text{tr}^2}_{n}}+\widehat{w}^{\text{tr}^2}_{n} - P^{\max}_{\text{tr}}.
\end{align}

Further, to make problem \eqref{eq_optimization_P4_l} differentiable, $I_-(x)$ is approximated as:
\begin{equation}
\label{eq_I_-}
\widehat{I}_-(x)=-\frac{1}{t}\log(-x),
\end{equation}
where $t$ is a parameter that sets the accuracy of the approximation. The larger the $t$, the closer the $\widehat{I}_-(x)$ is to ${I}_-(x)$.

Consequently, for a specific $t$, the optimization problem \eqref{eq_optimization_P4_l}  becomes:

\begin{align}
\label{eq_optimization_barrier_approx}
\min_{\left\lbrace \overline{w}^{\text{tr}}_{n}\right\rbrace,~\left\lbrace \widehat{w}^{\text{tr}}_{n}\right\rbrace} \:\: &-z_{\text{DC}}^{(l)}(\left\lbrace \overline{w}^{\text{tr}}_{n}\right\rbrace,~\left\lbrace\widehat{w}^{\text{tr}}_{n}\right\rbrace)-\sum_{i=1}^{2}\frac{\log(-f_{c,i}(\left\lbrace \overline{w}^{\text{tr}}_{n}\right\rbrace,~\left\lbrace \widehat{w}^{\text{tr}}_{n}\right\rbrace))}{t},
\end{align}
which can be solved by GD methods such as Newton's Method. The whole algorithm, named as SCP-IP, is summarized in Algorithm \ref{SCP}.

 {\textit{Remark 3:} One concern with Model II is that the correspondent input signal incurs unlimited bandwidth (equivalent to the output signal of a non-linear inverse HPA feeding the transmit waveform) and might be impractical from the implementation aspect. However, if the  {out-of-band}  frequency components of the input signal decay rapidly, the desired input signal can be approximated by cutting off those insignificant frequency components.}

\begin{algorithm}[h]
 $\textbf{Input}$: $\left( \left\lbrace\overline{w}^{\text{tr}}_{n}\right\rbrace,~\left\lbrace\widehat{w}^{\text{tr}}_{n}\right\rbrace\right)^{(0)},\epsilon_0>0,l\leftarrow 1$\;\\
 $\textbf{Output}$: $\left( \left\lbrace\overline{w}^{\text{tr}}_{n}\right\rbrace,~\left\lbrace\widehat{w}^{\text{tr}}_{n}\right\rbrace\right)^{\star}$;\\
 $\textbf{Repeat}$: \\
$\:\:\:\:\:\:1: \:$Compute $ (\left\lbrace\overline{\alpha}_{n}\right\rbrace,~\left\lbrace\widehat{\alpha}_{n}\right\rbrace)^{(l)}  $   using Eq. \eqref{eq_alpha_1} and Eq. \eqref{eq_alpha_2} at the operating point $\left( \left\lbrace\overline{w}^{\text{tr}}_{n}\right\rbrace,~\left\lbrace\widehat{w}^{\text{tr}}_{n}\right\rbrace\right)^{(l-1)}$ using Taylor expansion;\\
$\:\:\:\:\:\:2: \text{Compute } \left( \left\lbrace\overline{w}^{\text{tr}}_{n}\right\rbrace,~\left\lbrace\widehat{w}^{\text{tr}}_{n}\right\rbrace\right)^{(l)}$ using Algorithm \ref{algorithm_barrier};\\
$\:\:\:\:\:\:3: \:$Update $\left( \left\lbrace\overline{w}^{\text{tr}}_{n}\right\rbrace,~\left\lbrace\widehat{w}^{\text{tr}}_{n}\right\rbrace\right)^{\star}\leftarrow \left( \left\lbrace\overline{w}^{\text{tr}}_{n}\right\rbrace,~\left\lbrace\widehat{w}^{\text{tr}}_{n}\right\rbrace\right)^{(l)}$;\\
$\:\:\:\:\:\:4: \:$Quit if $|z_{\text{DC}}\left( \left\lbrace\overline{w}^{\text{tr}}_{n}\right\rbrace,~\left\lbrace\widehat{w}^{\text{tr}}_{n}\right\rbrace\right)^{(l)}-z_{\text{DC}}\left( \left\lbrace\overline{w}^{\text{tr}}_{n}\right\rbrace,~\left\lbrace\widehat{w}^{\text{tr}}_{n}\right\rbrace\right)^{(l-1)}|< \epsilon_0$;\\
$\:\:\:\:\:\:\:5: \:l\leftarrow l+1$;\;
\caption{SCP-IP}
\label{SCP}
\end{algorithm}
\begin{algorithm}[h]
 $\textbf{Input}$: $\left( \left\lbrace\overline{w}^{\text{tr}}_{n}\right\rbrace,~\left\lbrace\widehat{w}^{\text{tr}}_{n}\right\rbrace\right)^{(B_0)}\leftarrow\left( \left\lbrace\overline{w}^{\text{tr}}_{n}\right\rbrace,~\left\lbrace\widehat{w}^{\text{tr}}_{n}\right\rbrace\right)^{(l-1)},\:t>0,$\\
 $\:\:\:\:\:\:\:\:\:\:\:\:\:\:\:\:\:\:\:\:\:\:\:\:\:\:\:\:\:\:\:\:\:\:\:\:\:\:\:\:\:\:\:\:\:\:\:\:\:\:\:\:\mu_B>0,\epsilon_B>0$;\\
 $\textbf{Output}$: $\left( \left\lbrace\overline{w}^{\text{tr}}_{n}\right\rbrace,~\left\lbrace\widehat{w}^{\text{tr}}_{n}\right\rbrace\right)^{(l)}$; \\
 $\textbf{Repeat}$: \\
 $\:\:\:\:\:\:\:1:\:$Compute $\left( \left\lbrace\overline{w}^{\text{tr}}_{n}\right\rbrace,~\left\lbrace\widehat{w}^{\text{tr}}_{n}\right\rbrace\right)$ by minimizing problem \eqref{eq_optimization_barrier_approx} using Newton's Method with initialized point $\left( \left\lbrace\overline{w}^{\text{tr}}_{n}\right\rbrace,~\left\lbrace\widehat{w}^{\text{tr}}_{n}\right\rbrace\right)^{(B_0)}$;\\
 $\:\:\:\:\:\:\:2:\text{Update }\left( \left\lbrace\overline{w}^{\text{tr}}_{n}\right\rbrace,~\left\lbrace\widehat{w}^{\text{tr}}_{n}\right\rbrace\right)^{(l)}\leftarrow\left( \left\lbrace\overline{w}^{\text{tr}}_{n}\right\rbrace,~\left\lbrace\widehat{w}^{\text{tr}}_{n}\right\rbrace\right)$;\\
 $\:\:\:\:\:\:\:3: \:\text{Quit if } 2/t < \epsilon_B$;\\
$\:\:\:\:\:\:\:4: \:t\leftarrow\mu_Bt,\:\left( \left\lbrace\overline{w}^{\text{tr}}_{n}\right\rbrace,~\left\lbrace\widehat{w}^{\text{tr}}_{n}\right\rbrace\right)^{(B_0)}\leftarrow\left( \left\lbrace\overline{w}^{\text{tr}}_{n}\right\rbrace,~\left\lbrace\widehat{w}^{\text{tr}}_{n}\right\rbrace\right)^{(l)}$;\;
\caption{IP}
\label{algorithm_barrier}
\end{algorithm}

\textit{Remark 4:} From the mechanisms of SCP-SQP and SCP-IP, we can deduce the pros and cons of these two algorithms. For SCP-SQP, one of its limitations comes from the rapidly increasing number of variables and non-linear constraints given large $N$. Due to its origination from the KKT-NR method, this algorithm might be sensitive to bad performance, especially with large-scale problems (numerous constraints are involved), and will rely on the initialisation point heavily. In contrast, the SCP-IP strategy is only constrained by two inequalities and formulates a convex sub-program in each iteration, which provides more robustness. However, its integral non-linear constraint makes the sub-program more time-consuming on the other hand. {Unfortunately, it is hard to mathematically evaluate the complexity of the algorithms, especially for the SCP-IP in Model II where an integral inequality constraint is involved. However, numerous simulations show a polynomial time-complexity with the increase  {in} the number of variables, as well as a linear rate of convergence for the SCP for both algorithms. As expected, the SCP-IP from Model II requires a longer time due to its non-linear inequality constraint but shows a tendency to be more efficient given a large number of variables \footnote{During optimization, Model II takes advantage of the convexity of HPA's inverse characteristics, forming a convex problem finally. In contrast, for Model I, we adopt a very general and easily extended algorithm, SCP-SQP. Other methods have also been tried, but all involve much higher complexity or severely degraded performance. For instance, complexity increases exponentially when using the block coordinate descend method based on a popular polynomial approximation of HPA, where the amplitude is solved by geometric programming and the phase is solved by line search. The complexity mainly comes from HPA's non-linearities, which  {require} a relatively high polynomial order to grasp its saturation characteristics.  Also, the contradictory effect of HPA's and EH's non-linearities dissolves the most beneficial features that can be utilized to transform and simplify the optimization formulation, such as a monotonic relationship between the objective/constraint functions and potential optimization variables.  Severe performance degradation incurs when purely adopting SQP/IP or when simply adapting the input waveform to match the optimal transmit waveform assuming an ideal HPA, following a bisection search for an achievable power budget.}\footnote{ {Both models and their corresponding optimizations can be extended to MISO scenarios in a straightforward way. This paper focuses on SISO to avoid potential tedious subscripts. However, a MISO simulation is provided in Section V as a verification.}}.}

\section{Simulations}
\label{section_simulations}
The power harvesting and  {the end-to-end }PTE performance of the proposed waveforms are evaluated under a Wi-Fi-like scenario with $f_0=5.18$ GHz and bandwidth $B=10$ MHz unless specified.  We set a $38$ dBm transmit power, a $2$ dBi receive antenna gain and a $58$ dB path loss. For the simulation baseline of SSPA, set the smoothing parameter to $\beta=4$,  {the saturation voltage to $A_s=10$ dBV} and the small-signal gain to $G=1$ without loss of generality.

\subsection{Frequency-flat channels}
\label{simu_ff}
We start from frequency-flat channels to evaluate the power harvesting and  {the end-to-end PTE} performance of WPT  and to uncover the waveform tendency considering both HPA's and EH's non-linearit {ies}. The frequency-flat scenario is defined as $\widetilde{h}_{n}=1$ for $\forall~n$.
\begin{figure}[t]
\centering
\includegraphics[width=0.46\textwidth]{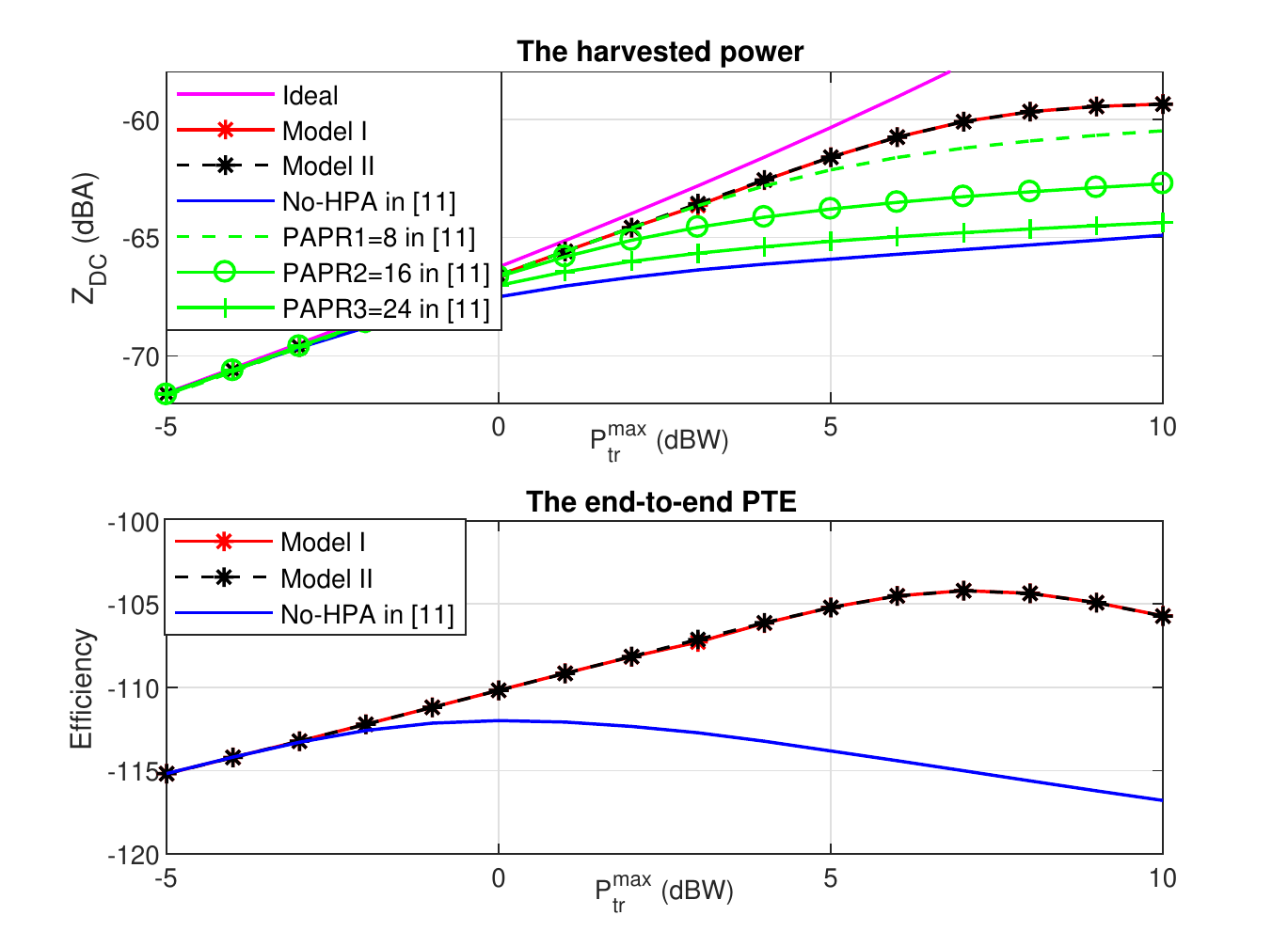}
\caption{$z_{\text{DC}}$ when considering both HPA and rectenna's non-linearity, $G=1, ~As=10~$dBV, $\beta=4,~N=8,~B=10$ MHz. Ideal refers to the maximal $z_{\text{DC}}$ with an ideal HPA. Model I/II refers to the resultant waveform optimized from Model I/II; No-HPA in \cite{Clerckx2016Waveform} is to use the  {optimal transmit} waveform from \cite{Clerckx2016Waveform} directly to HPA's input. PAPR is to use the waveform from \cite{Clerckx2016Waveform} but with PAPR constraints.}
\label{fig4}
\end{figure}

\begin{figure}[t]
\centering
\includegraphics[width=0.46\textwidth]{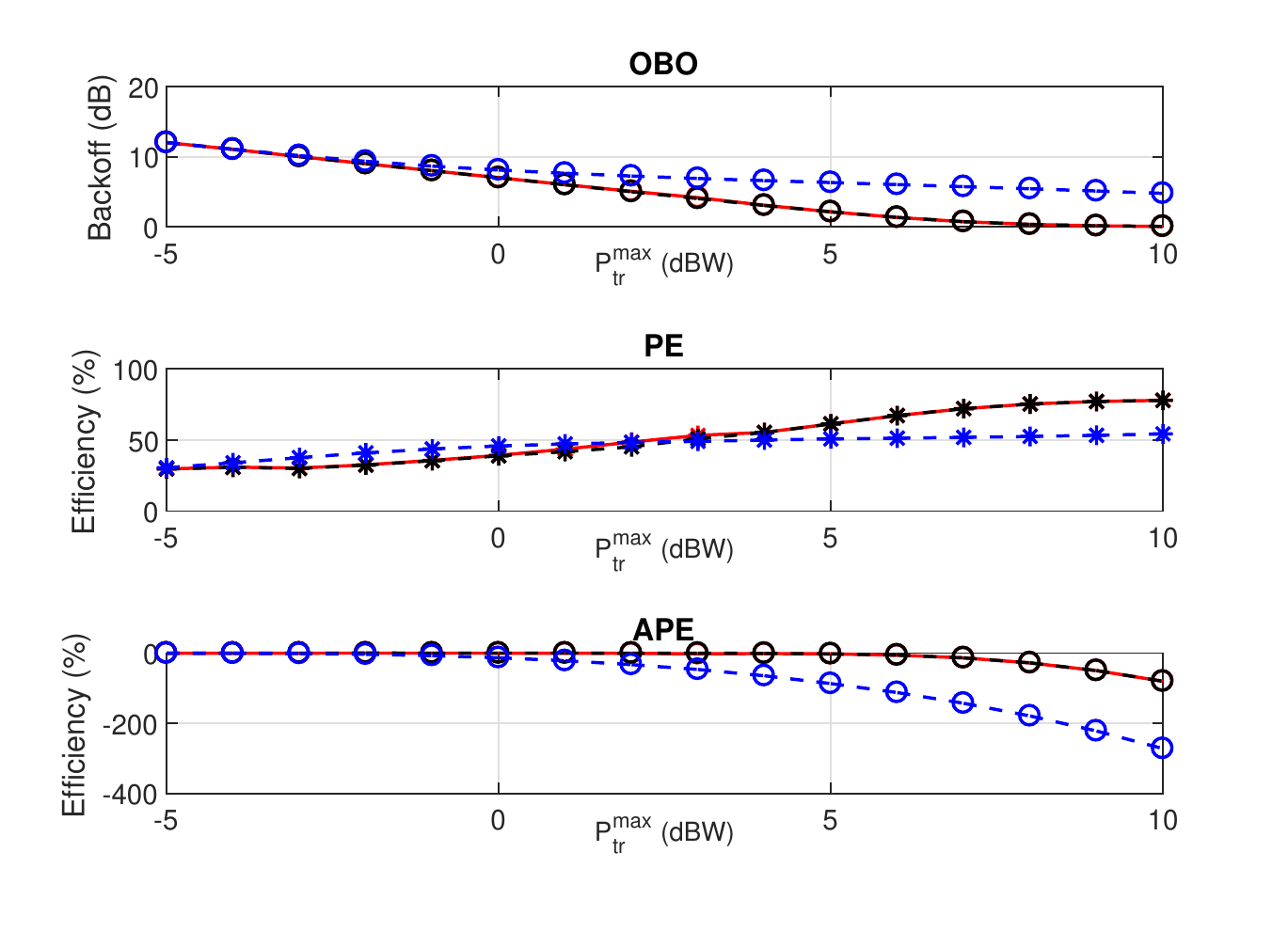}
\caption{The operating point of HPA that corresponds to Fig. \ref{fig4}. The red, black and blue curves represent the waveform from Model I, Model II and No-HPA in [11] respectively.}
\label{fig4_Amplifier}
\end{figure}
\begin{figure}[t]
\centering
\includegraphics[width=0.46\textwidth]{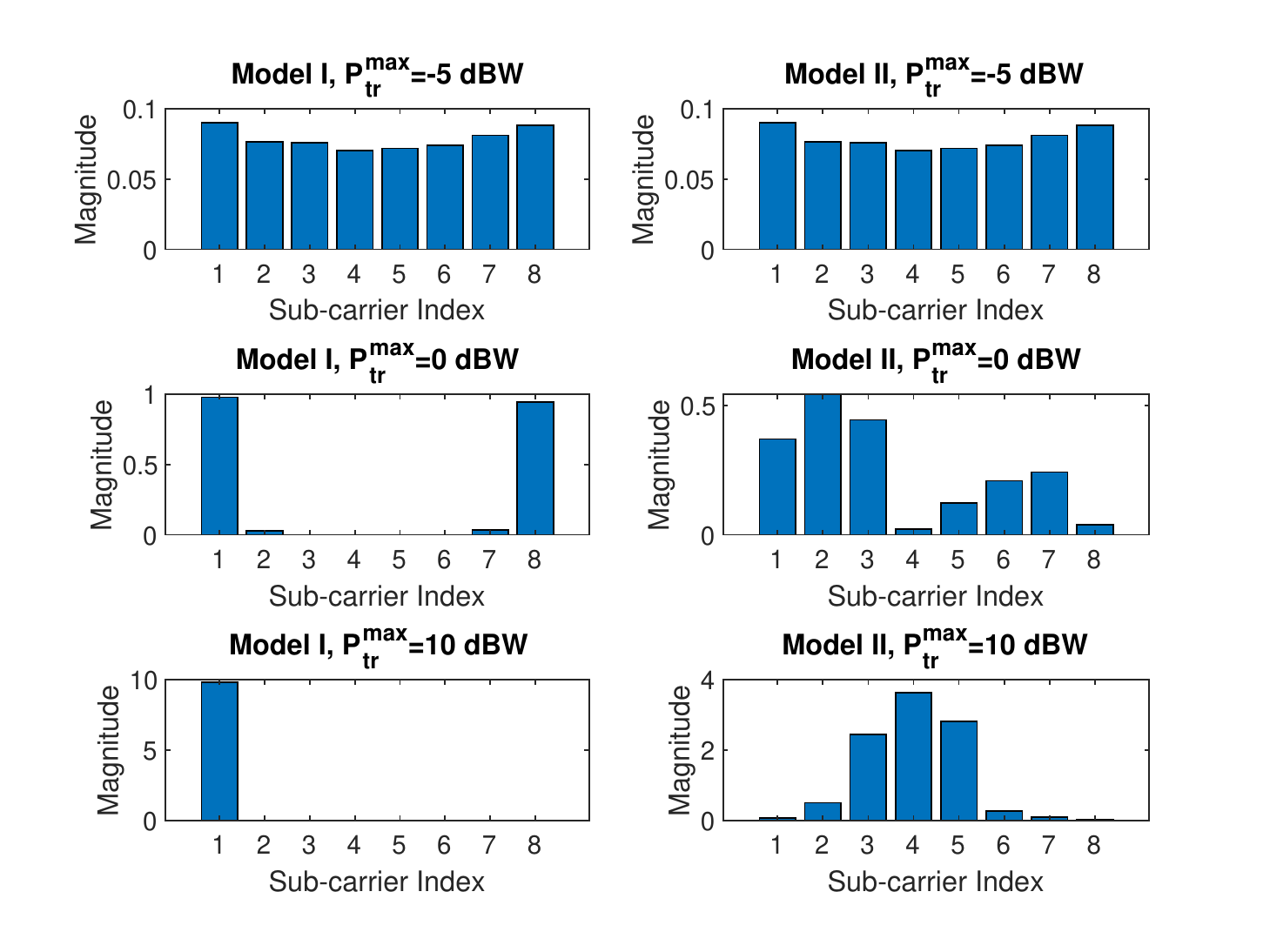}
\caption{The waveforms from the two end-to-end  {optimizations that correspond} to Fig. \ref{fig4} and Fig. \ref{fig4_Amplifier}, with HPA working in small signal regime, non-linear regime and saturation regime  {from top to the bottom}.}
\label{fig4_2}
\end{figure}

Fig. \ref{fig4} plots ${z}_{DC}$ (top) and the end-to-end PTE (bottom) in a frequency-flat WPT system as a function of the input/transmit power constraint ($P^{\max}_{\text{in}}=P^{\max}_{\text{tr}}$) using different input waveforms at the HPA, where ${z}_{DC}$ of an ideal amplifier is used (the  {magenta} line) as a benchmark. Fig. \ref{fig4} (top) demonstrates that HPA's non-linearity significantly reduces the harvested power in WPT, especially when the transmit power increases to a level comparable with HPA's saturation power. On the other hand, the harvested power (Fig. \ref{fig4}, top) verifies the performance gain of the two end-to-end optimized waveforms (Model I, red, and Model II, black) over the optimal transmit waveform in \cite{Clerckx2016Waveform} (the blue line) as the input of the non-linear HPA. Nevertheless, the transmit waveform with PAPR constraints (green) can compensate for HPA's non-linearity to a certain degree. {As a response to Remark 3, the performance and analysis of an approximated input signal of Model II is given in Appendix D. }

To gain a deeper insight into HPA's non-linear effect on the power harvesting performance and the waveform tendency, Fig. \ref{fig4} (bottom) plots the end-to-end PTE  {defined in} Eq. \eqref{eq_E2E_PTE}.  Fig. \ref{fig4_Amplifier} evaluates HPA's operating points (severity of non-linearity and PE) by characterizing its OBO, PE, and APE that correspond to the simulation points in Fig. \ref{fig4}. Fig. \ref{fig4_2} then plots three examples of the optimal input waveforms from Model I and Model II,  {respectively with HPA working in its linear, non-linear and saturation regime}. Details are as follows.

In the small-signal region ($P^{\max}_{\text{tr}}<-3$ dBW, OBO $>10$ dB), the HPA works in the linear regime and can be regarded as ideal, i.e. negligible power loss or out-of band frequency leakage. In this case, the waveforms obtained from Model I and Model II in Fig. \ref{fig4_2} (top) overlap, both giving the maximal harvested power the same as that from the optimal transmit waveform with ideal HPA. However, in this region, the PE of HPA is low ($<40\%$) as shown in  Fig. \ref{fig4_Amplifier} (middle). Increasing the input power exploits the HPA's DC power supply better, and hence gives larger PE (Fig. \ref{fig4_Amplifier}, middle), APE (Fig. \ref{fig4_Amplifier}, bottom) and also the end-to-end PTE in WPT (Fig. \ref{fig4} bottom).

However, when the transmit power increases to over $-3$ dBW ( OBO $<10$ dB), the HPA's non-linear effect becomes non-negligible. In this non-linear region ($-3$ dBW $\leq P^{\max}_{\text{tr}}\leq 8$ dBW), while the harvested power still increases with  {the transmit power}, the performance gap between different waveforms enlarges (Fig. \ref{fig4}). The waveforms from Model I and Model II give the best performance, followed by using the optimal transmit waveform with and without PAPR constraints. The reason for the performance gap is indicated in Fig. \ref{fig4_Amplifier} (top), where the two end-to-end optimization waveforms provide significantly smaller OBO, or equivalently speaking, larger output power after HPA than the No-HPA waveform in \cite{Clerckx2016Waveform}. This is achieved by allocating power to fewer sub-carriers as shown in Fig. \ref{fig4_2} (middle).  Such an allocation exploits HPA's DC power supply better and is proved effective to compensate for HPA's non-linear degradation in Fig. \ref{fig4_Amplifier} (bottom), where the HPA's APE of the proposed input waveforms of HPA (Model I and Model II) maintains the same level as the small signal regime. Consequently, we still observe an increase in the end-to-end PTE in Fig. \ref{fig4} (bottom) for the proposed waveforms, whereas the end-to-end PTE of the No-HPA waveform starts decreasing because its high-PAPR waveform comes across severe power loss from HPA's non-linear degradation effect. 

 {Interestingly, although in Fig. \ref{fig4}, the performance of Model I and Model II almost overlap in the non-linear region, their corresponding optimal waveforms in Fig. \ref{fig4_2} (middle) are quite different from each other. Taking the point at $ P^{\max}_{\text{tr}}=0$ dBW as an example, the transmit waveform from Model I shows slightly higher PAPR than that from Model II ($5.07$ over $4.33$) but lower transmit power ($-0.01$ dBW over $-0.02$ dBW) on the other hand. The former makes the waveform from Model I beneficial for the EH's non-linearity, while the latter indicates that the waveform from Model II passes through the HPA and BPF with less power loss, resulting in their similar power harvesting performance in the end. Such a waveform difference is attributed to their signal models. The waveform from Model II tends to have a relatively lower PAPR transmit solution because its maximal  {magnitude of the transmit signal, equivalent to the signal after HPA,} is strictly restricted by the HPA's saturation voltage. A lower PAPR (and also the lossless BPF) makes the waveform from Model II more beneficial from the transmitter's side. In comparison, the waveform from Model I caters to the EH's non-linearity slightly better.
}

If further increasing the power constraint to $P^{\max}_{\text{Tr}}>8$ dBW, the HPA's saturation effect will dominate, in which condition the harvested power tends to saturate rather than increase (Fig. \ref{fig4}). The resultant optimal input waveforms (Fig. \ref{fig4_2}, bottom) show a further reduction in  PAPR. Despite this reduction in PAPR, Fig. \ref{fig4_Amplifier} (bottom) illustrates that waveform design can no longer compensate for HPA's non-linear degradation, giving APE lower than 0. Understandably, in this region, although the harvested power (Fig. \ref{fig4}) and the HPA's PE (Fig. \ref{fig4_Amplifier}, middle) keep non-decreasing, the end-to-end PTE of all the waveforms drops with increasing transmit power (Fig. \ref{fig4} bottom).  { {This end-to-end PTE trend indicates inappropriate operating points in WPT for $P^{\max}_{\text{tr}}>8$ dBW.} Inspired by this, most of the following simulations are based on the  {operating point that gives} the best overall performance in terms of both $z_{\text{DC}}$ and the end-to-end PTE, i.e., $P^{\max}_{\text{tr}}=8$ dBW for $A_s=10$ dBW and $\beta=4$.} 

\begin{figure}[!t]
\centering
\includegraphics[width=0.46\textwidth]{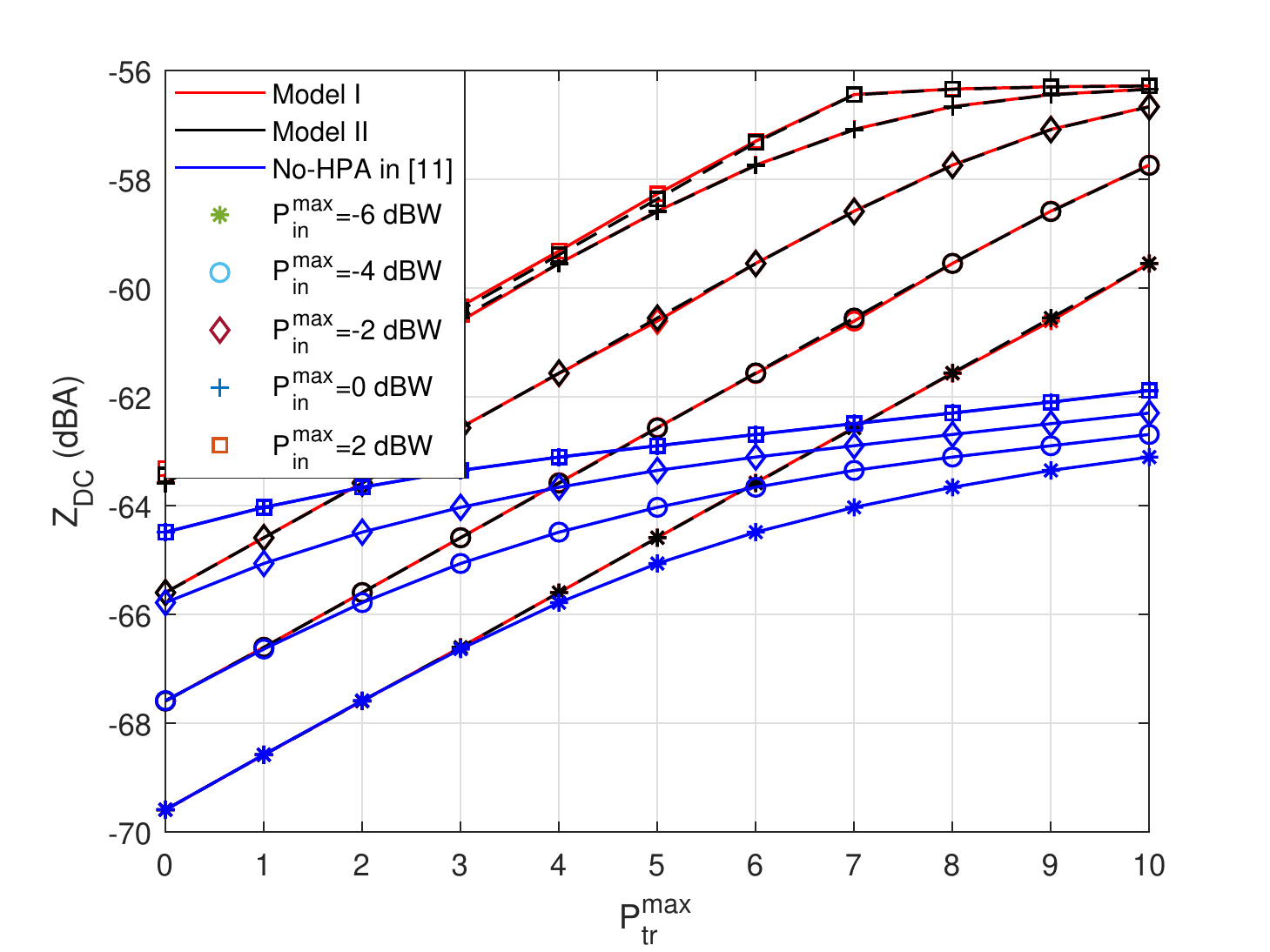}%
\caption{$z_{\text{DC}}$ of different waveforms in frequency-flat channels with different levels of the input power constraints as a function of the  {transmit} power constraints, $G=1,~As=10~$dBA, $\beta=4,~ N=8,~B=10$MHz.}
\label{fig5}
\end{figure}

An extension of Fig. \ref{fig4} is shown in Fig. \ref{fig5}, where different levels of input power are compared ($P^{\max}_{\text{in}}=P^{\max}_{\text{in}}+\left\lbrace-6,~-4,~-2,~0,~2\right\rbrace$ dBW) in terms of the power harvesting performance and the end-to-end PTE. An indication here is that increasing the input power can compensate for HPA's non-linear degradation to some extent (Fig. \ref{fig5}). This is reasonable since larger input power gives larger transmit power at the antenna and therefore, larger harvested power at the receiver. However, this does not necessarily give higher end-to-end  {PTE} since a larger input signal might suffer more severe power loss due to HPA's non-linearity. 

\subsection{Frequency-selective channels}
We then evaluate the power harvesting performance of different waveforms in frequency-selective scenarios. We assume a large open space environment with an NLOS channel from model B in \cite{etsi1998channel} with 18 delay taps whose power is normalized to 1. For each realization, we assume that the complex channel gain for each tap is modelled as an independent circularly symmetric complex Gaussian random variable, with zero mean and the variance being their normalized tap power.
\begin{figure*}[!t]
\centering
\subfloat[]{\includegraphics[width=0.46\textwidth]{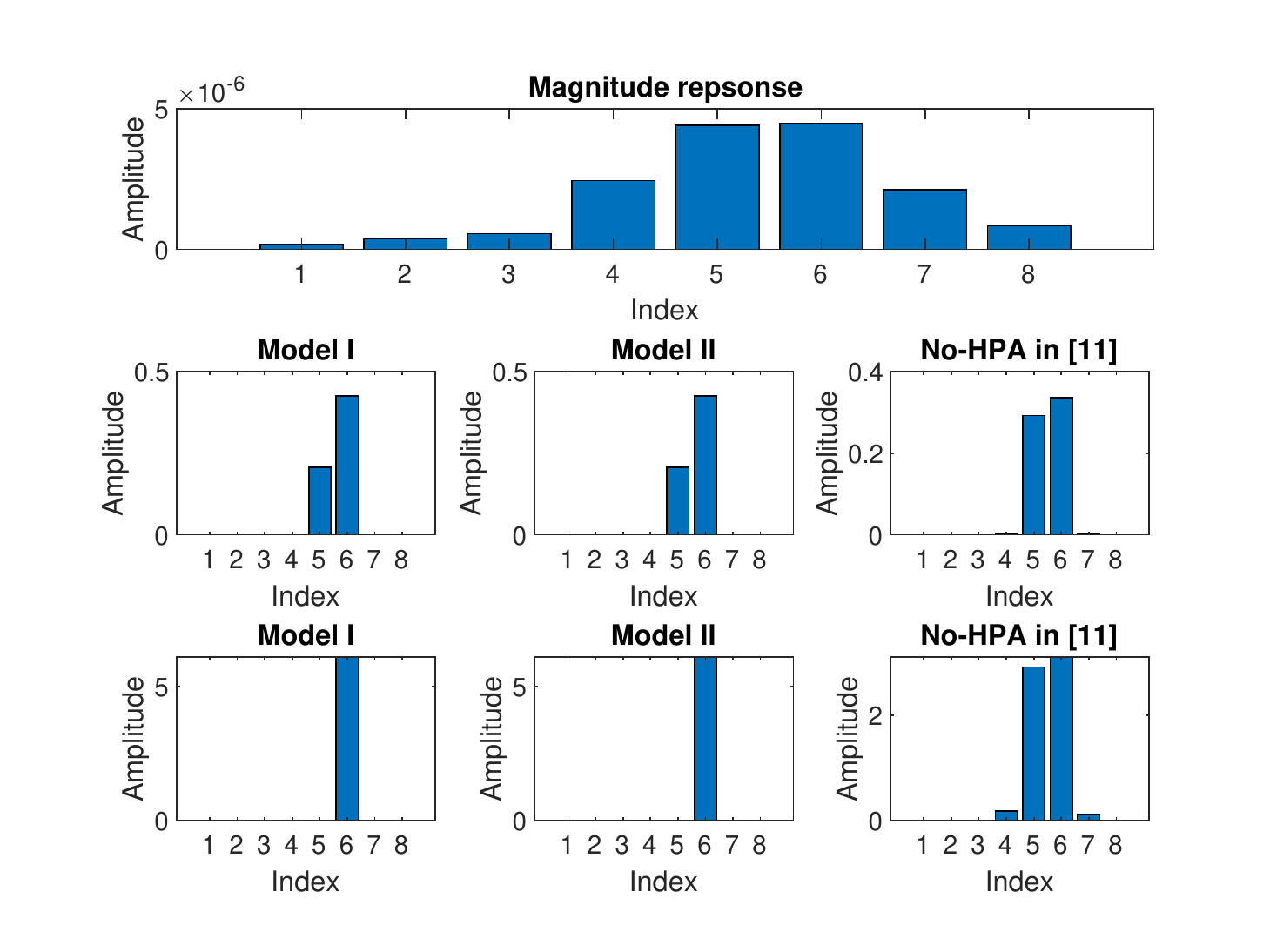}%
\label{fig4_selective_CSI}}
\hfil
\subfloat[]{\includegraphics[width=0.46\textwidth]{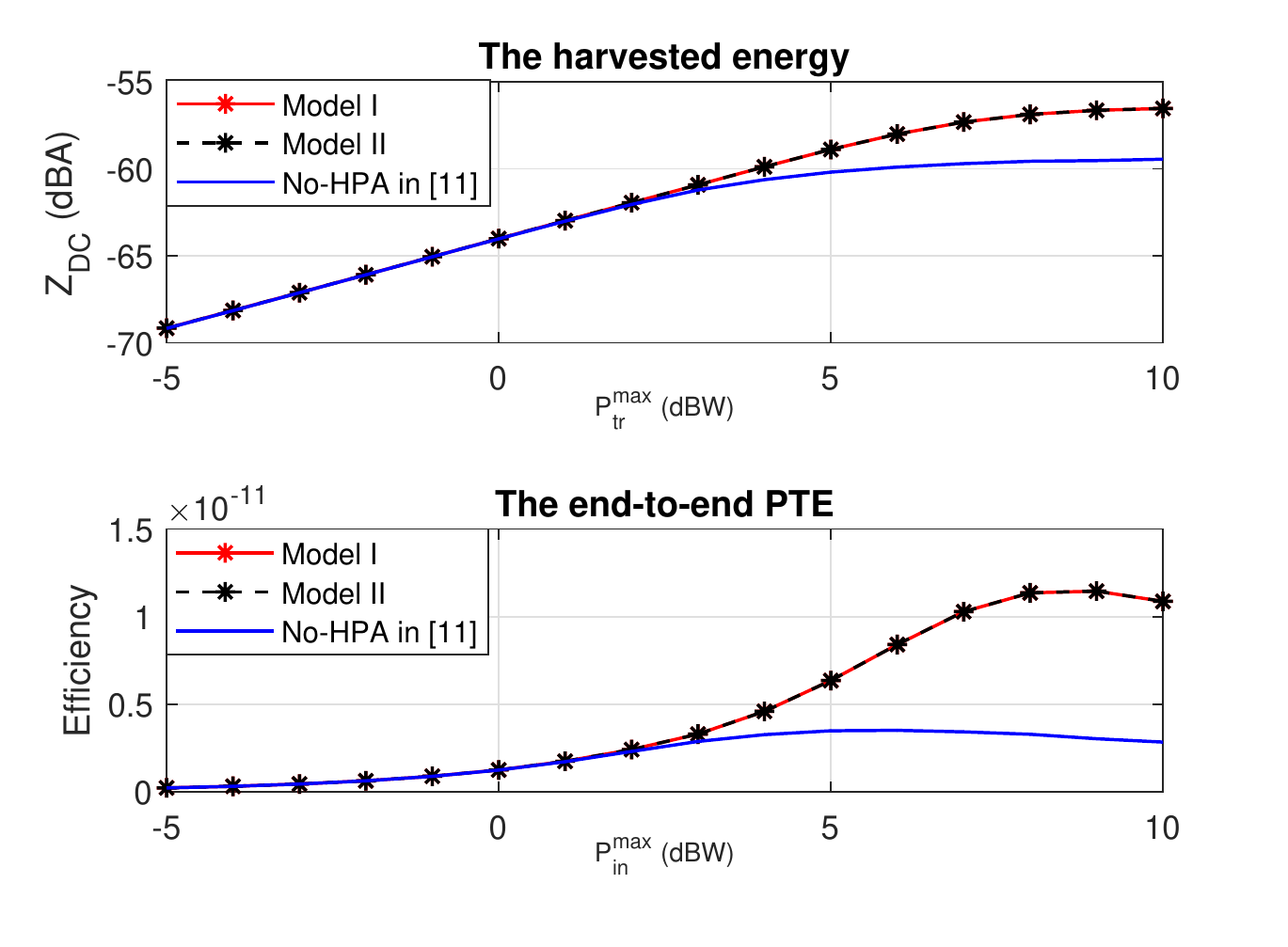}%
\label{fig4_selective_EH}}
\caption{The optimal waveforms and their power harvesting performance given a frequency-selective channel realisation, $G=1,~As=10~$dBV, $\beta=4,~N=8,\:B=10$MHz.  Fig. \ref{fig4_selective_CSI}, from top to bottom, plots the magnitude response of this channel realisation, the waveform from the three models with $P^{\max}_{\text{tr}}=-5$dBW, and the waveform from the three models with $P^{\max}_{\text{tr}}=5$dBW. Fig. \ref{fig4_selective_EH} plots the harvested power $z_{\text{DC}}$ on the top and the end-to-end PTE at the bottom.}
\label{fig4_selective}
\end{figure*}

As a comparison to the frequency-flat scenario, Fig. \ref{fig4_selective_CSI} and Fig .\ref{fig4_selective_EH} evaluate the waveforms' performance based on one realization of a frequency-selective channel. Fig. \ref{fig4_selective_CSI} (top) illustrates the magnitude response of the channel realization\footnote{We only explore the effect of CSI's magnitudes here, since the phases of channels do not affect the power harvesting performance significantly from simulations.}. Fig. \ref{fig4_selective_CSI} (middle and bottom) displays the magnitude of the optimal waveforms from Model I, Model II and the No-HPA waveform in \cite{Clerckx2016Waveform} in linear and non-linear region ($P^{\max}_{\text{in}}=-5/5$ dBW) respectively.  Fig. \ref{fig4_selective_EH} plots the corresponding harvested power (top) and the end-to-end PTE (bottom).

Fig. \ref{fig4_selective_CSI} (middle and bottom) shows that the No-HPA waveform also features relatively  {lower} PAPR in this frequency-selective channel realization. As a result, a smaller performance gap between the proposed waveforms and the No-HPA waveform is  {observed} in Fig. \ref{fig4_selective_EH}, compared with that in Fig. \ref{fig4}. Also interestingly, with HPA's non-linearity, Fig. \ref{fig4_selective_EH} (top) displays larger harvested power than that from the frequency-flat scenario in Fig. \ref{fig4}. This is reasonable since, with significant HPA's non-linearity, the proposed waveforms from Model I and Model II concentrate power on sub-carriers with the strongest channels for both frequency-flat and frequency-selective scenarios. On the other hand, the normalized frequency-selective channels offer more powerful strongest sub-carriers than the normalized frequency-flat channels. Consequently, the end-to-end PTE (Fig.  \ref{fig4_selective_EH}, bottom) with frequency-selective channels is also significantly larger than that of the frequency-flat channels in Fig. \ref{fig4} (bottom).

 {Compared with the optimal waveforms from the frequency-flat channels in Fig. \ref{fig4_2}, the optimal waveforms from the frequency-selective channels in Fig. \ref{fig4_selective_CSI} do not show much difference. This is because the EH's non-linearity is preferring a lower PAPR waveform in frequency-selective scenarios, and mitigates the trade-off between the HPA's and  the EH's non-linearity.
}

\begin{figure}[t]
\centering
\includegraphics[width=0.46\textwidth]{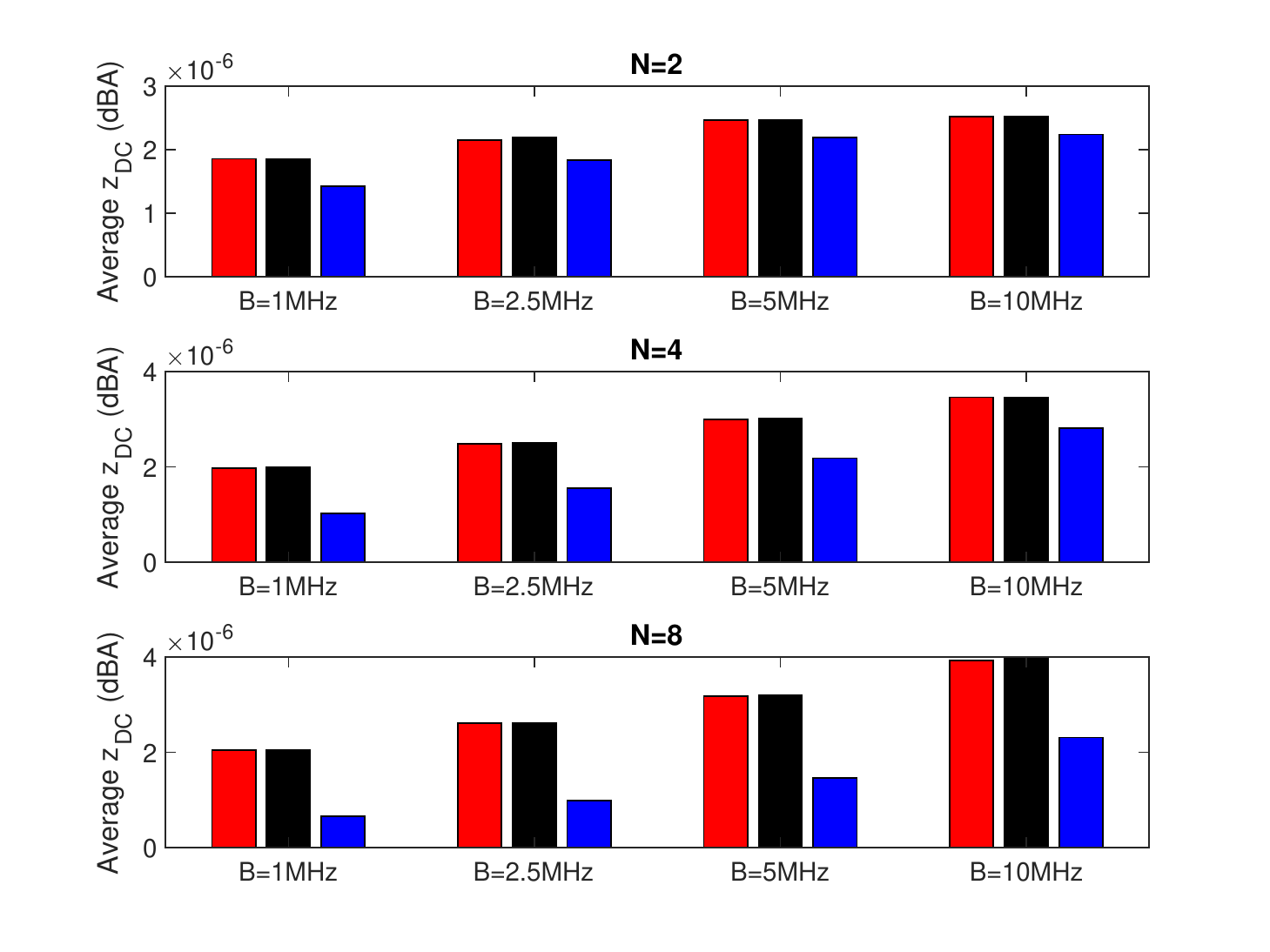}
\caption{Average $z_{\text{DC}}$ with different bandwidth, $P^{\max}_{\text{In}}=10$ dBW, $P^{\max}_{\text{Tr}}=8$ dBW, $G=1,~ As=10~$dBW, $\beta=4$. The red and black bars represent the waveforms from Model I and Model II respectively. The blue bar represents the No HPA waveform in [11].}
\label{fig7}
\end{figure}
Fig. \ref{fig7} further identifies the effect of the bandwidth on the average harvested power over hundreds of channel realizations. It demonstrates that all these three waveforms benefit from larger bandwidth, attributing to their common preference for strong channel transmission. Larger bandwidth provides more diversity between channels and thus has a larger probability of owning a more powerful strongest sub-carrier. This is highly beneficial especially when HPA's non-linearity urges single-carrier transmission. Even for the No-HPA waveform in [11] that does not consider HPA's non-linearity, transmission with larger bandwidth gives rise to lower PAPR solutions of the No-HPA waveform and mitigates its power loss through the non-linear HPA.

\begin{figure}[t]
\centering
\includegraphics[width=0.46\textwidth]{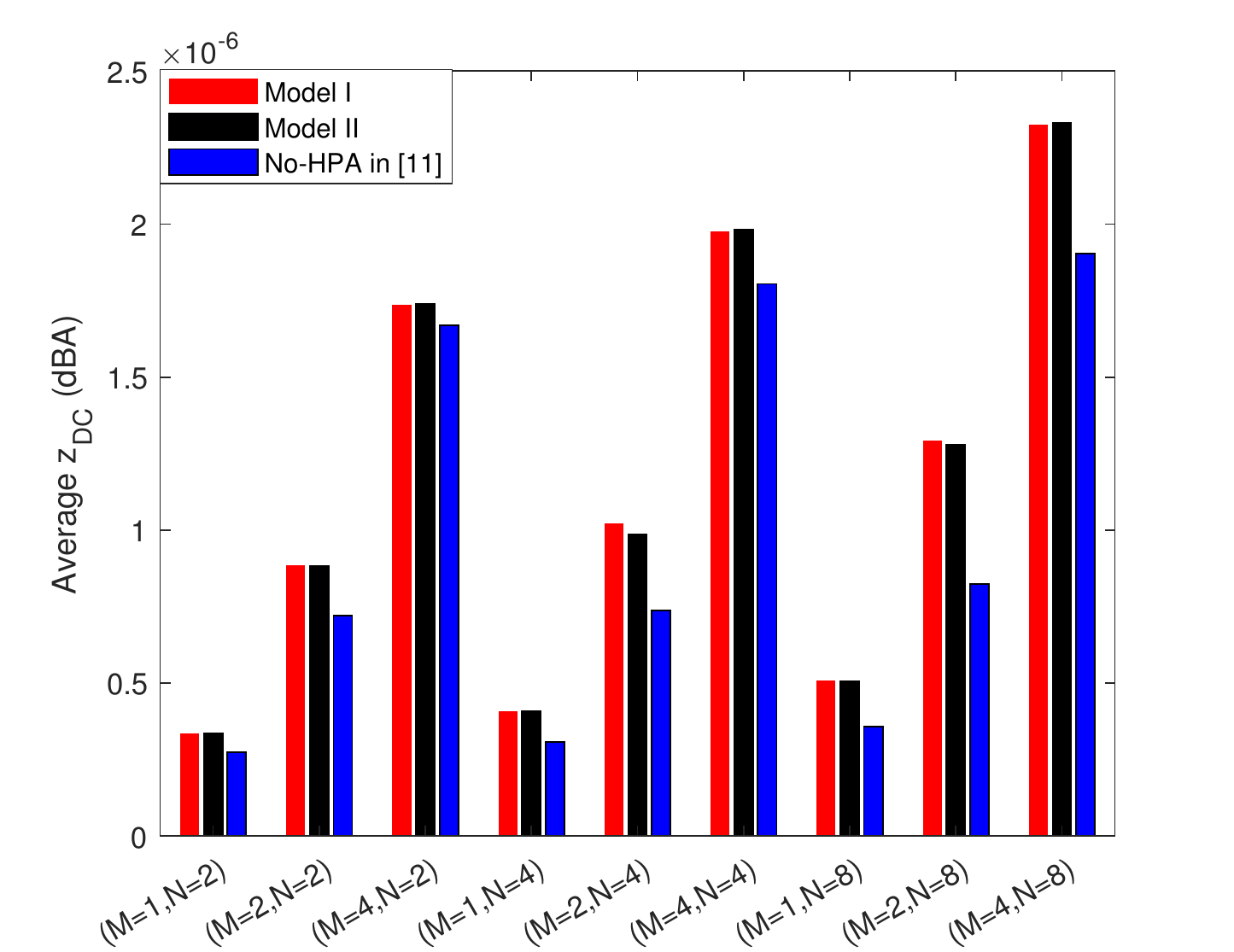}
\caption{ {Average $z_{\text{DC}}$ with different numbers of antennas $M$ and different numbers of sub-carriers $N$ for $P^{\max}_{\text{In}}=10$ dBW, $P^{\max}_{\text{Tr}}=8$ dBW, $G=1, As=10~$dBA, $\beta=4$. The red and black bars represent the waveforms from Model I and Model II respectively. The blue bar represent the No HPA waveform in [11].}}
\label{fig7_diff_M}
\end{figure}
 Although we omit MISO in the previous formulations, the models in Section III and the optimizations in Section IV can be straightforwardly extended to MISO scenarios, whose average harvested power is plotted in  Fig. \ref{fig7_diff_M}.  Fig. \ref{fig7_diff_M} shows that increasing the number of antennas benefits both the performance of Model I and Model II, since it provides more channel diversity. However, in contrast with Fig. \ref{fig7} where the No-HPA waveform also benefits from the increased bandwidth, in Fig. \ref{fig7_diff_M}, the performance gap between the No-HPA waveform and the proposed  {waveforms} enlarges with the increasing number of antennas. This originates from the 'matched filter' solution of the No-HPA waveform in MISO, where power is allocated to each antenna proportionally to the channel strength, making the strongest antenna more vulnerable to heavier power loss from HPA. On the contrary, the waveforms from Model I and Model II can obtain a flexible trade-off between exploiting channel's diversity and reducing HPA's power loss.

\begin{figure}[t]
\centering
\includegraphics[width=0.46\textwidth]{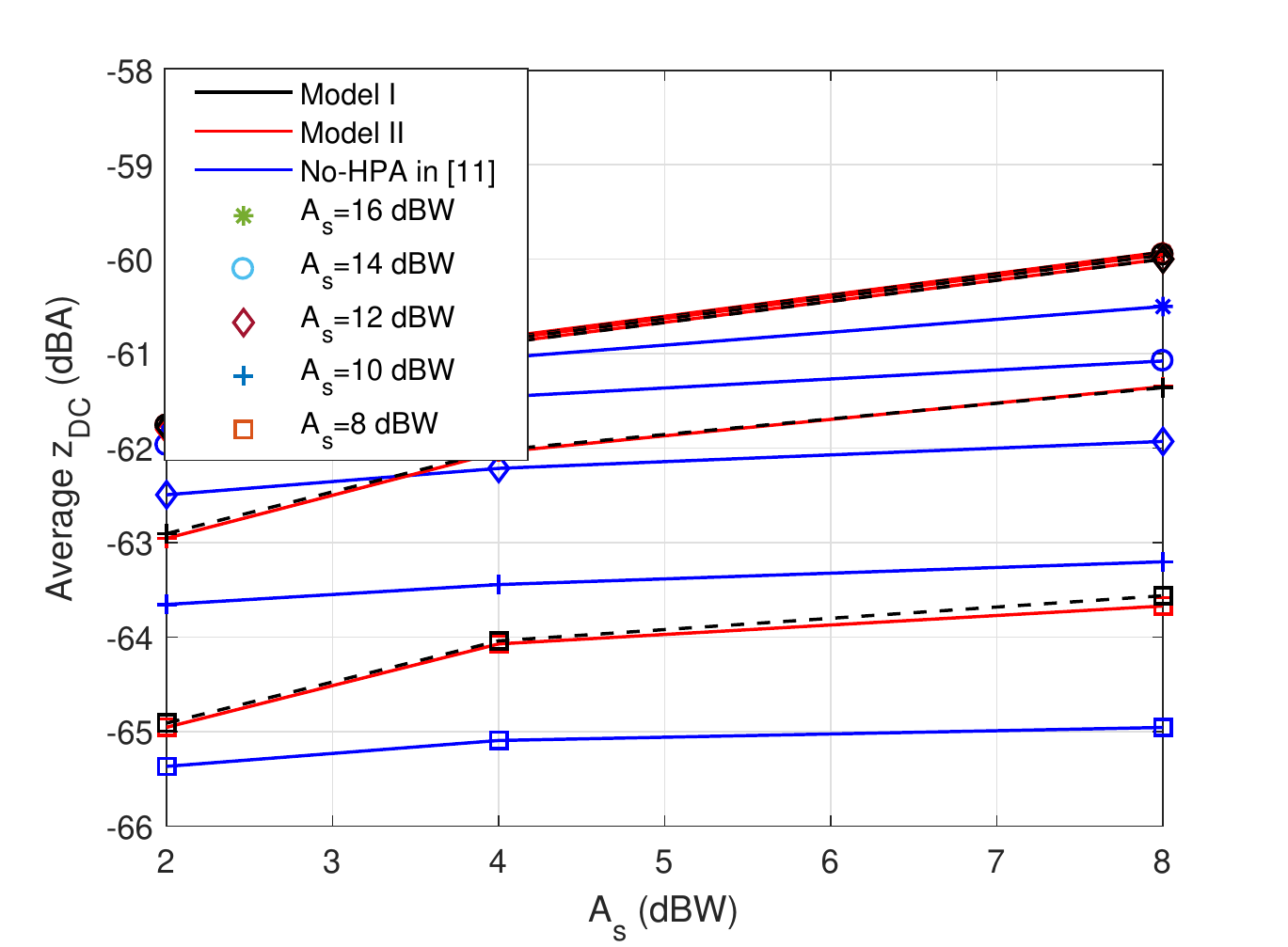}
\caption{Average $z_{\text{DC}}$ with different numbers of sub-carriers, giving different saturation power, $P^{\max}_{\text{In}}=10$, $P^{\max}_{\text{Tr}}=8$ dBW, $G=1,~ \beta=4,~B=10~$MHz.}
\label{fig6}
\end{figure}
Fig. \ref{fig6} compares the average harvested power as a function of the number of sub-carriers with different saturation voltages of HPA. In \cite{Clerckx2016Waveform}, without HPA's non-linearity, the ideal harvested power increases proportionally to the number of sub-carriers. In contrast, in Fig.  \ref{fig6}, the average harvested power shows a tendency to saturate when increasing the number of sub-carriers. This comes from a fine balance between HPA's non-linearity (favours low-PAPR transmission) and EH's non-linearity (favours relative high-PAPR transmission), which consequently favours only a limited number of sub-carriers for transmission. The more severe the HPA's non-linearity, the more dominant HPA's non-linear effect exerts on the input waveform design. Correspondingly, in Fig.  \ref{fig6}, smaller saturation power indicates a higher non-linear operating point of HPA (smaller OBO) and thus shows a more significant saturation tendency as well as less harvested power.

\subsection{ {Comparison between Model I and Model II}}
\begin{figure}[t]
\centering
\includegraphics[width=0.46\textwidth]{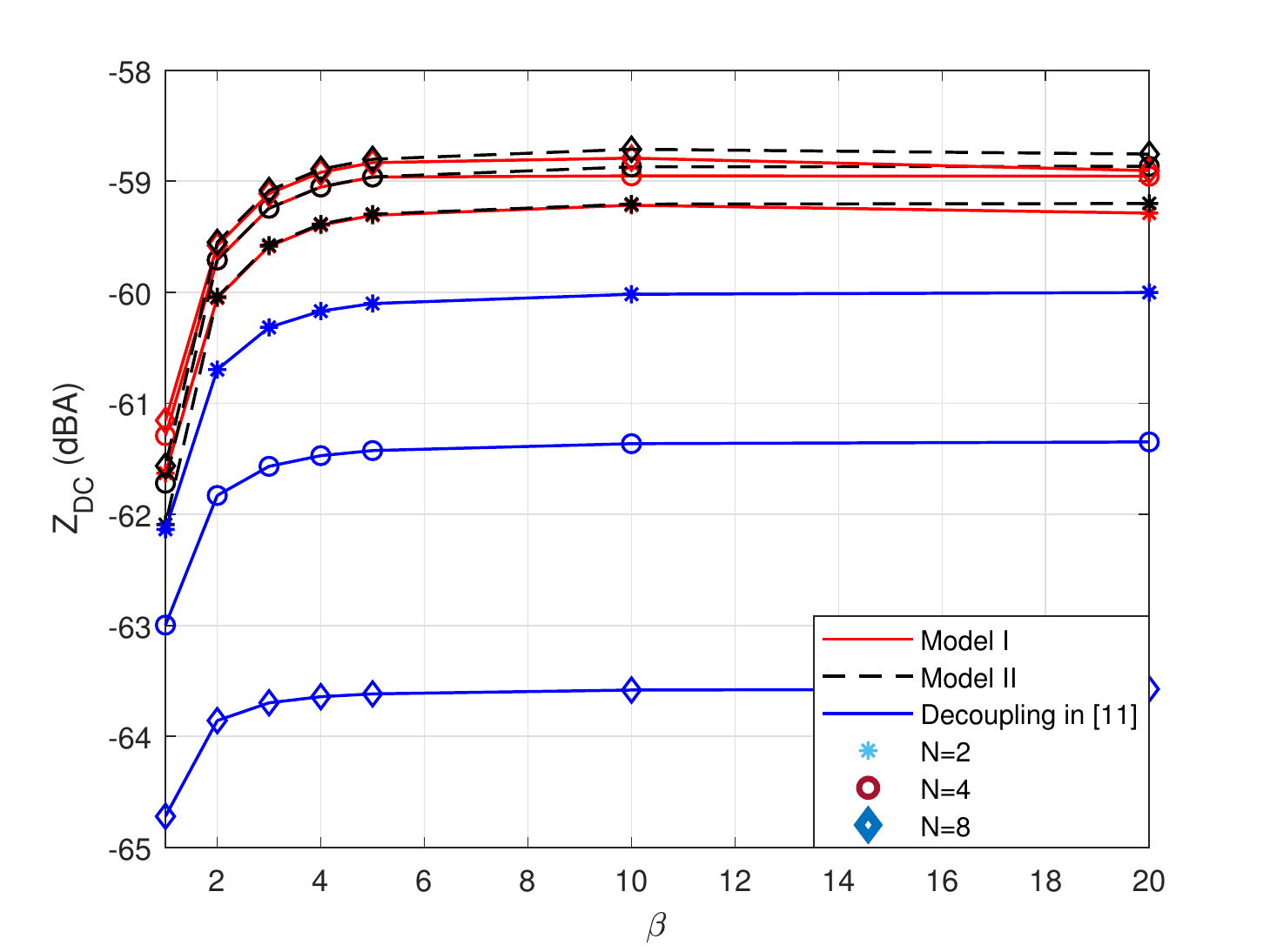}
\caption{Average $z_{\text{DC}}$ as a function of HPA's smoothing parameter ($\beta$) given different number of sub-carriers, $P^{\max}_{\text{In}}=10$ dBW, $P^{\max}_{\text{Tr}}=8$ dBW, $G=1,~A_s=10$ dBA, $B=10~$MHz.}
\label{fig9}
\end{figure}
 So far, no significant difference has been observed between the two end-to-end optimized waveforms from Model I and Model II. As an explanation, first, the waveforms from Model I and Model II converge to very similar solutions in the small-signal linear region and the saturation region. For the former, they approach the No-HPA waveform since the EH's non-linear effect dominates, while for the latter, they approach single-carrier transmission to mitigate the power loss caused by the HPA's non-linearity. Second, even within the non-linear region where the two models give different waveforms as in Fig. \ref{fig4_2} (middle), the performance gap between Model I and Model II in Fig. \ref{fig4} is still not significant. For one reason, as it has been mentioned in section  \ref{simu_ff}, the waveform from Model I features relatively higher PAPR and is more advantageous for the EH's non-linearity while the waveform from Model II suffers less power loss at the transmitter, finally leading to a similar end-to-end power harvesting performance between Model I and Model II. For another reason, the previous simulations are based on an HPA's smoothing parameter of $\beta=4$, which is commonly used but whose transfer characteristic experiences a sharp saturation after the linear region (Fig. \ref{fig_SSPA_model}). Compared with $\beta=1$, a higher smoothing parameter of $\beta=4$ decreases the performance gap between the two models because most of the signal can be regarded as working in a linear region.

 {Hence, to reveal the performance difference between Model I and Model II, we plot the power harvesting performance of different HPAs, characterized by different smoothing parameters in Fig. \ref{fig9}. } Fig. \ref{fig9} firstly shows that HPA with a smaller smoothing parameter produces less harvested power. This is reasonable since, from Fig. \ref{fig_SSPA_model}, the HPA with a smaller smoothing parameter shows higher non-linearity over the operation regime and thus suffers more severe power loss for the same input signals.

 {Also, Fig. \ref{fig9} indicates that given a small smoothing parameter (i.e., $\beta=1$ where HPA suffers significant non-linearity even below the saturation voltage in Fig. \ref{fig_SSPA_model}), the waveform from Model I has slightly better power harvesting performance than Model II, whereas for HPA with large smoothing parameters (i.e., $\beta=10$ where HPA is highly linear below its saturation voltage in Fig. \ref{fig_SSPA_model}), the waveform from Model II outperforms slightly.  {Their different preferences on HPA's smoothing parameters still originate from the models.} {As mentioned in section \ref{simu_ff}, Model II is more advantageous in obtaining higher transmit power by   {generating a relatively lower PAPR waveform}, which not only mitigates HPA's power loss but also makes the waveform free from the {power} loss from the BPF.} However, with a small smoothing parameter, the high non-linearity of HPA forces both models to generate a low PAPR waveform and pass the BPF in an almost lossless way. { In this case, the transmit power advantage of Model II diminishes and the PAPR advantage of Model I becomes more significant. On the contrary, when increasing the smoothing parameter, Model I is generating higher PAPR waveforms and is experiencing increasing power loss from the BPF. When the smoothing parameter becomes  large enough (i.e., $\beta=10$ for example), Model I is outperformed by Model II because its power loss from BPF becomes non-negligible.}}

\section{Conclusions}
This paper investigates the optimal input waveform of HPA that maximizes the power harvesting performance in WPT, accounting for both HPA's and rectenna's non-linearities and adaptive to the CSI.  Two models are proposed and solved based on whether the input waveform of HPA has the flexibility to transmit over the frequencies out of the transmit pass band or not. Simulations verify the advantages of these two  optimized input waveforms at HPA, especially with highly non-linear HPA and in frequency-flat channels. The proposed waveforms show that, as the HPA  {experiences} more non-linearity, the optimal waveforms transform from high-PAPR to low-PAPR or even to single-carrier transmission. This reveals the trade-off between HPA's and rectenna's non-linearity, i.e., the former causes power loss and can be compensated by low-PAPR transmission, while the latter boosts the harvested power and is better exploited with high-PAPR transmission. We also show that while increasing the input signal power makes use of HPA's DC power supply better and generates higher harvested power, the end-to-end PTE of the WPT system decreases when operating in HPA's highly non-linear (saturation) regime, since HPA's non-linear degradation becomes more and more severe.  {We also compare the waveforms from the two models and show that they share very similar performance in most scenarios, but they feature different waveforms and might slightly {outperform each other  {dependent} on HPA parameters}.}

\section{Appendix}
\subsection{Convexity proof of the objective function in Eq. \eqref{eq_optimization_P2_1}}
The objective function in Eq. \eqref{eq_optimization_P2_1} is easily seen to be increasing and convex with respect to $ \overline{y}(t)^2=\overline{y}( \left\lbrace\overline{w}^{\text{tr}}_{n}\right\rbrace,~\left\lbrace\widehat{w}^{\text{tr}}_{n}\right\rbrace,~t)^2$  {from Eq. \eqref{eq_scaling_term0}}.  {On the other hand, if omitting $t$ here, $\overline{y}\left( \left\lbrace\overline{w}^{\text{tr}}_{n}\right\rbrace,~\left\lbrace\widehat{w}^{\text{tr}}_{n}\right\rbrace\right)^2\overset{\Delta}{=}y_1(y_2( \left\lbrace\overline{w}^{\text{tr}}_{n},~\widehat{w}^{\text{tr}}_{n}\right\rbrace))$ is a composition of a convex function $y_1(x)=x^2$ and an affine function $y_2( \left\lbrace\overline{w}^{\text{tr}}_{n},~\widehat{w}^{\text{tr}}_{n}\right\rbrace)=\sum_{n=0}^{N-1}\left(\overline{h}_n\cos(2\pi f_n t)-\widehat{h}_n\sin(2\pi f_n t)\right)\overline{w}^{\text{tr}}_{n}-\left(\widehat{h}_n\cos(2\pi f_n t)+\overline{h}_n\sin(2\pi f_n t)\right)\widehat{w}^{\text{tr}}_{n}$ with respect to the joint variables $ \left\lbrace\overline{w}^{\text{tr}}_{n},~\widehat{w}^{\text{tr}}_{n}\right\rbrace $. Such a composition makes $\overline{y}( \left\lbrace\overline{w}^{\text{tr}}_{n},~\widehat{w}^{\text{tr}}_{n}\right\rbrace)^2$ convex with respect  to $ \left\lbrace\overline{w}^{\text{tr}}_{n},~\widehat{w}^{\text{tr}}_{n}\right\rbrace$ jointly. As a proof, }
\begin{equation}
\label{eq_composition}
 \overline{y}( \left\lbrace\overline{w}^{\text{tr}}_{n},~\widehat{w}^{\text{tr}}_{n}\right\rbrace)^2=y_1(y_2( \left\lbrace\overline{w}^{\text{tr}}_{n},~\widehat{w}^{\text{tr}}_{n}\right\rbrace)),
\end{equation}

 {Thus, we have:}
\begin{align}
\label{eq_g_convex}
& y_1(y_2(\theta \left\lbrace\overline{w}^{\text{tr}}_{n},~\widehat{w}^{\text{tr}}_{n}\right\rbrace_1+(1-\theta) \left\lbrace\overline{w}^{\text{tr}}_{n},~\widehat{w}^{\text{tr}}_{n}\right\rbrace_2))
\nonumber\\ & =y_1(\theta y_2( \left\lbrace\overline{w}^{\text{tr}}_{n},~\widehat{w}^{\text{tr}}_{n}\right\rbrace_1)+(1-\theta)y_2( \left\lbrace\overline{w}^{\text{tr}}_{n},~\widehat{w}^{\text{tr}}_{n}\right\rbrace_2))\\
&\leq \theta y_1( y_2( \left\lbrace\overline{w}^{\text{tr}}_{n}, \widehat{w}^{\text{tr}}_{n}\right\rbrace_1))+(1-\theta)f_1(y_2( \left\lbrace\overline{w}^{\text{tr}}_{n}, \widehat{w}^{\text{tr}}_{n}\right\rbrace_2)),
\end{align}
 {which provides the convexity.}

 {Hence, the objective $z_{\text{DC}}$ is increasing and convex with respect to $\overline{y}( \left\lbrace\overline{w}^{\text{tr}}_{n}\right\rbrace,~\left\lbrace\widehat{w}^{\text{tr}}_{n}\right\rbrace ,t)^2$ while $\overline{y}( \left\lbrace\overline{w}^{\text{tr}}_{n}\right\rbrace,~\left\lbrace\widehat{w}^{\text{tr}}_{n}\right\rbrace ,t)^2$ is convex with respect to $ \left\lbrace\overline{w}^{\text{tr}}_{n},   \widehat{w}^{\text{tr}}_{n}\right\rbrace$, which implies that $z_{\text{DC}}$ is convex with respect to $ \left\lbrace\overline{w}^{\text{tr}}_{n},  \widehat{w}^{\text{tr}}_{n}\right\rbrace$.}

\subsection{Proof of the input power constraint in Eq. \eqref{eq_optimization_P3_3}}
Firstly, we derive the inverse transfer characteristics of HPA (SSPA in Eq.\eqref{eq_HPA_model}). Given the amplitude distortion function $\mathcal{A}(\cdot)$ in Eq.\eqref{eq_HPA2_amp} and the phase shift $\Phi(\cdot)$ in Eq. \eqref{eq_HPA2_phase}, the inverse transfer function of HPA is characterized by $f^{-1}(\cdot)_{\text{HPA}}$, composed of the inverse amplitude function $\mathcal{A}^{-1}(\cdot)$ and the inverse phase shift function $\Phi^{-1}(\cdot)$ such that:
\begin{align}
\label{eq_inverse_condition}
f_{\text{HPA}}(f^{-1}_{\text{HPA}}(\widetilde{x}^{\text{tr}}))=\mathcal{A}(\mathcal{A}^{-1}(x^{\text{tr}}))e^{j(\measuredangle \widetilde{x}^{\text{tr}}+\Phi(x^{\text{tr}})+\Phi^{-1}(x^{\text{tr}}))}=\widetilde{x}^{\text{tr}},
\end{align}
where $\widetilde{x}^{\text{tr}}$/$ {x}^{\text{tr}}$ is short for $ \widetilde{x}^{\text{tr}}(t)$/$ {x}^{\text{tr}}(t)$, and $ \widetilde{x}^{\text{tr}}=x^{\text{tr}}e^{j\measuredangle \widetilde{x}^{\text{tr}}}$.

Hence, we obtain the inverse transfer characteristics of HPA:
\begin{align}
\label{eq_inverse_SSPA_AM/AM}
&\mathcal{A}^{-1}(x^{\text{tr}})=\frac{x^{\text{tr}}}{G}\left[{1-(\frac{{x}^{\text{tr}}}{A_s})^{2\beta}}\right]^{-\frac{1}{2\beta}}, \\
\label{eq_inverse_SSPA_AM/PM}
&\Phi^{-1}({x}^{\text{tr}})=0,\\
\label{eq_inverse_SSPA}
&f^{-1}_{\text{HPA}}(\widetilde{x}^{\text{tr}})=\frac{\widetilde{x}^{\text{tr}}}{G}\left[{1-(\frac{{x}^{\text{tr}}}{A_s})^{2\beta}}\right]^{-\frac{1}{2\beta}}.
\end{align}

Then, from Parseval's theorem (the signal power in the frequency domain equals the signal power in the time domain), combing Eq. \eqref{eq_inverse_SSPA}, we have:
\begin{align}
\label{eq_inverse_P_tr}
\sum_{n=0}^{N-1} \widetilde{w}^{\text{in}^2}_{n} =\frac{1}{T}\int_{T}|f^{-1}_{\text{HPA}}(\widetilde{x}^{\text{tr}})|^2 dt=\frac{1}{2TG^2}\int_{T}\left[{{x^{\text{tr}}}^{-2\beta}-A_s^{-2\beta}}\right]^{-\frac{1}{\beta}}dt,
\end{align}
where $\widetilde{x}^{\text{tr}}$ is short for the transmit signal, with ${x}^{\text{tr}}$ being the corresponding amplitude. Eq.\eqref{eq_inverse_P_tr} corresponds to the constraint in Eq. \eqref{eq_optimization_P3_3}.

\subsection{Convexity proof of the constraint in Eq. \eqref{eq_optimization_P3_3}}
Denote the constraint in Eq. \eqref{eq_optimization_P3_3} as $ f( \left\lbrace\overline{w}^{\text{tr}}_{n},~\widehat{w}^{\text{tr}}_{n}\right\rbrace)=h(g( \left\lbrace\overline{w}^{\text{tr}}_{n},~\widehat{w}^{\text{tr}}_{n}\right\rbrace))$, with $h(x)= \frac{1}{2TG^2}\int_{T}[{x^{-\beta}-A_s^{-2\beta}}]^{-\frac{1}{\beta}}dt$  and $ g( \left\lbrace\overline{w}^{\text{tr}}_{n},~\widehat{w}^{\text{tr}}_{n}\right\rbrace)=x^{\text{tr}^2}( \left\lbrace\overline{w}^{\text{tr}}_{n},~\widehat{w}^{\text{tr}}_{n}\right\rbrace,t)=|\sum_n^N (\overline{w}^{\text{tr}}_{n}+j\widehat{w}^{\text{tr}}_{n})e^{jw_nt}|^2$.

For $h(x)$, we have:
\begin{align}
\label{eq_h_first_derivative}
\bigtriangledown h(x)
=&\frac{1}{2TG^2}\int_{T} [1-(\frac{x}{A_s^2})^{\beta}]^{-\frac{\beta+1}{\beta}}dt>0
\end{align}

And the second derivative:
\begin{align}
\label{eq_h_second_derivative}
\bigtriangledown^2 h(x)
=&\frac{1}{2TG^2}\int_{T} \frac{\beta+1}{A_s^{2\beta}}[1-(\frac{x}{A_s^2})^{\beta}]^{-\frac{2\beta+1}{\beta}}x^{\beta-1}dt>0.
\end{align}

At the same time, $ g( \left\lbrace\overline{w}^{\text{tr}}_{n}, \widehat{w}^{\text{tr}}_{n}\right\rbrace)$ can be proved convex with respect to $  \left\lbrace\overline{w}^{\text{tr}}_{n}, \widehat{w}^{\text{tr}}_{n}\right\rbrace$, since again, $  g( \left\lbrace\overline{w}^{\text{tr}}_{n}, \widehat{w}^{\text{tr}}_{n}\right\rbrace)$ is a composition of a convex norm function with an affine function (Appendix A). Given increasing and convex $h(x)$, together with convex $  g( \left\lbrace\overline{w}^{\text{tr}}_{n}, \widehat{w}^{\text{tr}}_{n}\right\rbrace)$, we can see that $f( \left\lbrace\overline{w}^{\text{tr}}_{n}, \widehat{w}^{\text{tr}}_{n}\right\rbrace)$ is convex regarding $ \left\lbrace\overline{w}^{\text{tr}}_{n}, \widehat{w}^{\text{tr}}_{n}\right\rbrace$.

\subsection{Approximating the input signal of Model II}

\begin{figure}[t]
\centering
\includegraphics[width=0.46\textwidth]{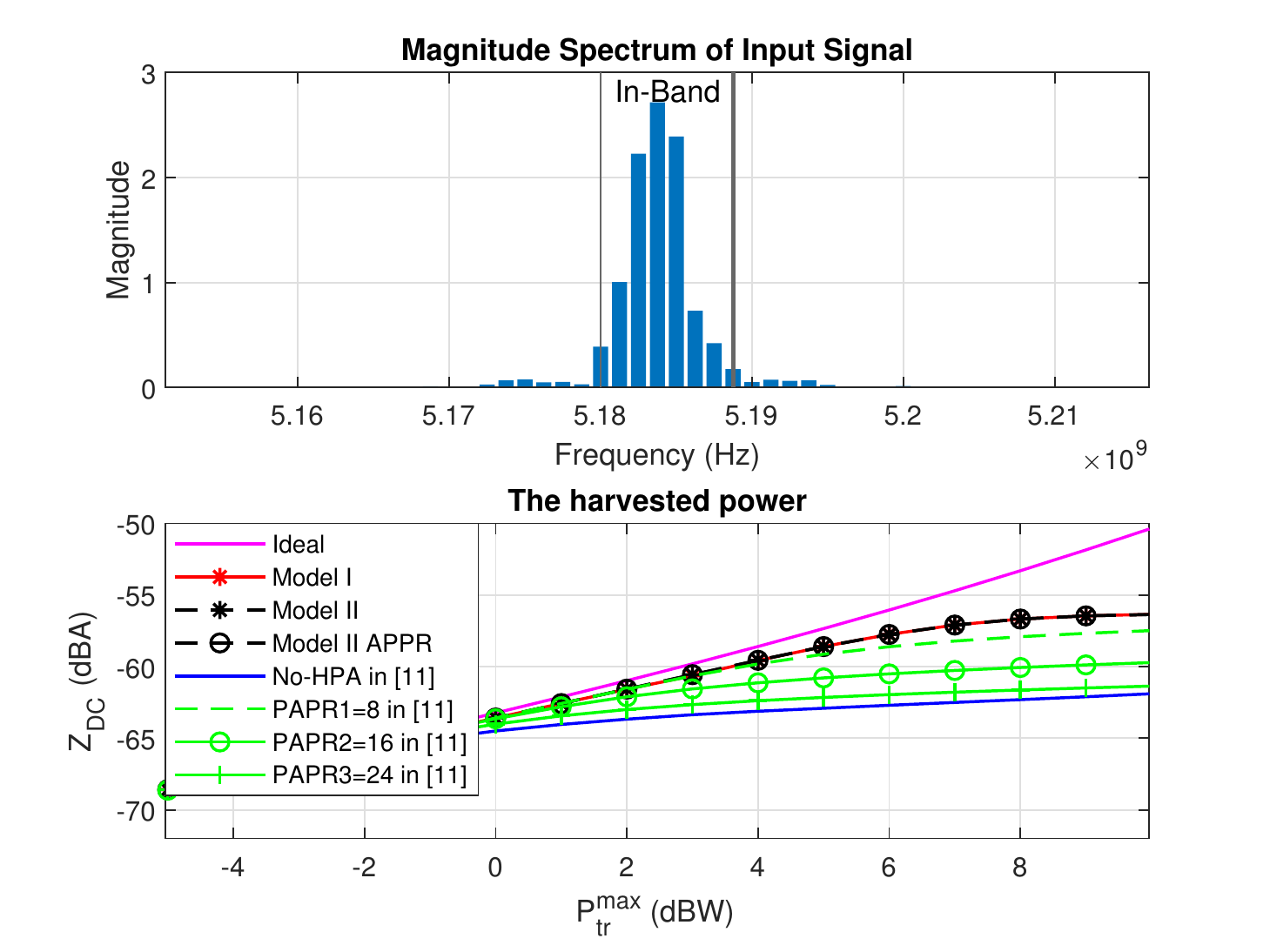}
\caption{ {The magnitude spectrum of the input signal from Model II (with $P^{\max}_{\text{tr}}=10$ dBW for the maximal non-linearity) (top) and the power harvesting performance corresponds to Fig. \ref{fig4} but adding an approximated input signal of Model II whose out-off band components are cut off (bottom).}}
\label{fig4_spectrum}
\end{figure}

 {Fig. \ref{fig4_spectrum} shows the spectrum of the input signal from Model II corresponding to Fig. \ref{fig4} on the top and plots the performance of an approximated signal (Model II APPR) whose  {out-of band} frequencies are cut off. It shows that the approximated signal achieves almost the same performance as the desired signal from the waveform 'Model II'. This can be explained by the magnitude spectrum on the top figure, where the  {out-of band} frequency components are almost negligible. Indeed, through simulations, even if introducing more HPA non-linearities by decreasing the smoothing parameter to $\beta=1$, the signal from 'Model II' can be perfectly approximated simply by extending the input bandwidth to $1.5$ times of the transmit bandwidth.}
\bibliographystyle{IEEEtran}
\bibliography{references.bib}

\begin{IEEEbiography}[{\includegraphics[width=1in,height=1.25in,clip,keepaspectratio]{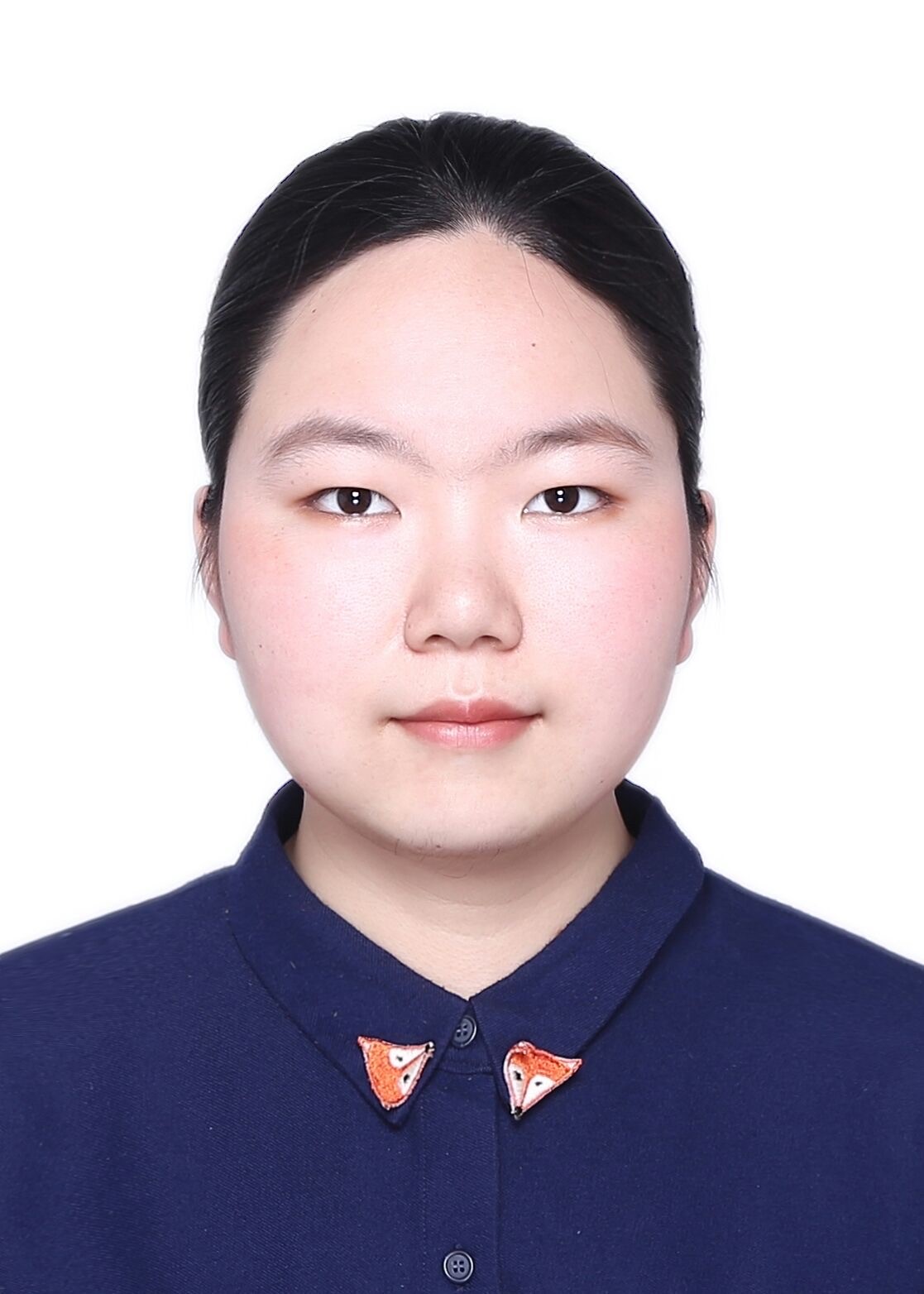}}]{Yumeng Zhang}
(Student Member, IEEE)
Yumeng Zhang received the B.S. degree in communication engineering from
the Nanjing University of Science and Technology, Nanjing, China, in 2019, and the M.S. degree in communications and signal processing from the Imperial College London, London, UK, in 2020. She is currently pursuing the Ph.D. degree with the Department of Electrical and Electronic Engineering, Imperial College London, U.K.
Her current research interests include signal processing, wireless communications, sensing and wireless power transfer.
\end{IEEEbiography}

\begin{IEEEbiography}[{\includegraphics[width=1in,height=1.25in,clip,keepaspectratio]{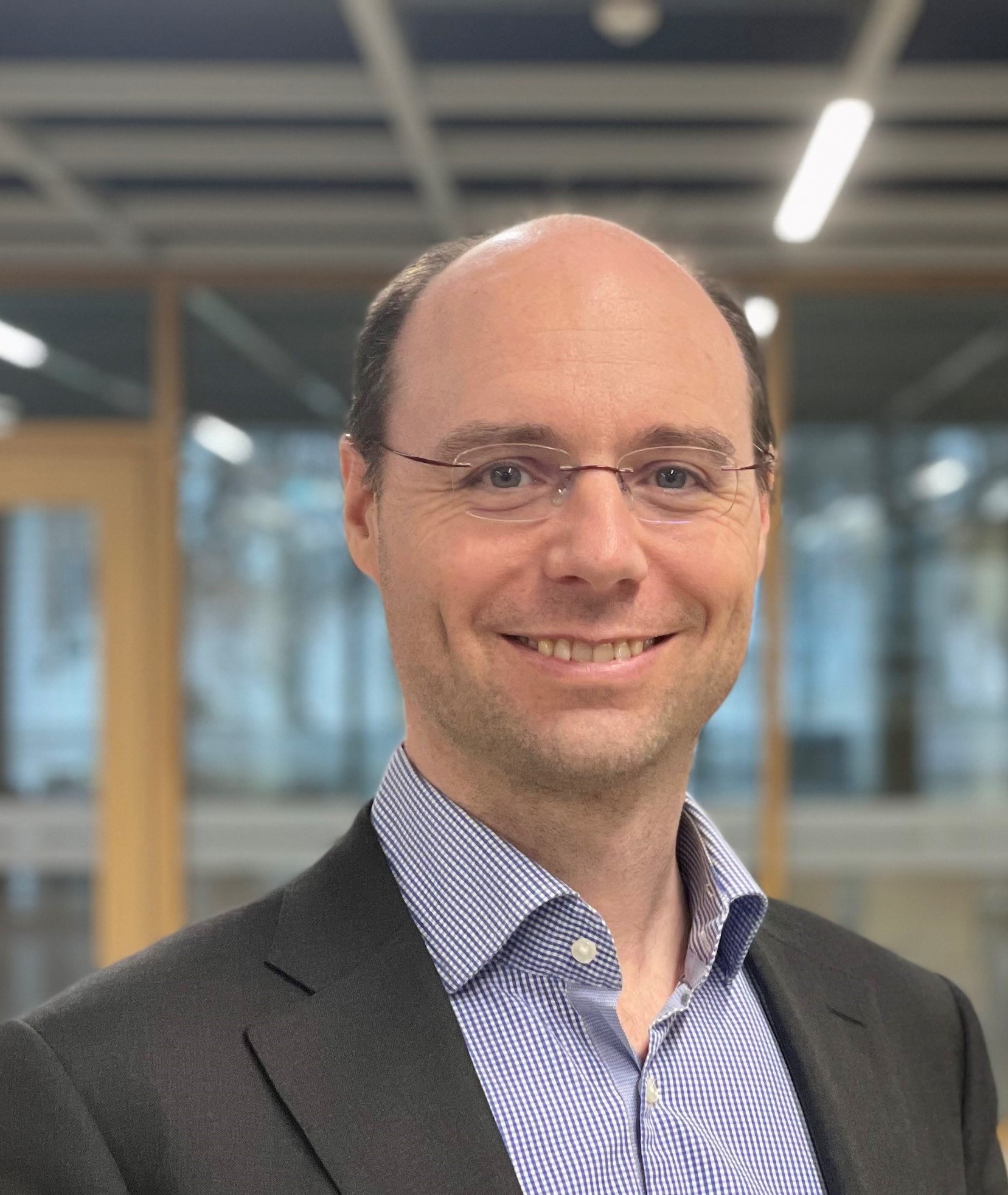}}]{Bruno Clerckx}
(Fellow, IEEE) Bruno Clerckx is a Professor, the Head of the Wireless Communications and Signal Processing Lab, and the Deputy Head of the Communications and Signal Processing Group, within the Electrical and Electronic Engineering Department, Imperial College London, London, U.K. He is the Chief Technology Officer (CTO) of Silicon Austria Labs (SAL) where he is responsible for all research areas of Austria's top research center for electronic based systems. He received the MSc and Ph.D. degrees in Electrical Engineering from Université Catholique de Louvain, Belgium, and the Doctor of Science (DSc) degree from Imperial College London, U.K. He worked for many years for Samsung Electronics, Korea, on 4G standardization (3GPP RAN1 and IEEE standards), and held various short- and long-term appointments at various institutions around the world, including Stanford University, Korea University, and Seoul National University. He has authored two books on “MIMO Wireless Communications” and “MIMO Wireless Networks”, 250 peer-reviewed international research papers, and 150 standards contributions, and is the inventor of 80 issued or pending patents. His research spans the general area of wireless communications and signal processing for wireless networks. He was an Elected Member of the IEEE Signal Processing Society “Signal Processing for Communications and Networking” (SPCOM) Technical Committee. He served as an Editor for the IEEE TRANSACTIONS ON COMMUNICATIONS, the IEEE TRANSACTIONS ON WIRELESS COMMUNICATIONS, and the IEEE TRANSACTIONS ON SIGNAL PROCESSING. He has also been a (lead) guest editor for special issues of the EURASIP Journal on Wireless Communications and Networking, IEEE ACCESS, the IEEE JOURNAL ON SELECTED AREAS IN COMMUNICATIONS, the IEEE JOURNAL OF SELECTED TOPICS IN SIGNAL PROCESSING, and the PROCEEDINGS OF THE IEEE. He was an Editor for the 3GPP LTE-Advanced Standard Technical Report on CoMP. He received the prestigious Blondel Medal 2021 from France for exceptional work contributing to the progress of Science and Electrical and Electronic Industries, the 2021 Adolphe Wetrems Prize in mathematical and physical sciences from Royal Academy of Belgium, multiple awards from Samsung, IEEE best student paper award, and the EURASIP (European Association for Signal Processing) best paper award 2022. He is a Fellow of the IEEE and the IET, and an IEEE Communications Society Distinguished Lecturer.
\end{IEEEbiography}

\end{document}